\documentclass[9pt,twocolumn,twoside]{opticajnl}
\journal{josab}
\doi{}
\setboolean{shortarticle}{false}
\setboolean{displaycopyright}{false}

\usepackage{amsmath,amssymb,bm,mathtools}
\usepackage{mathrsfs}
\usepackage{booktabs}
\usepackage{graphicx}
\usepackage{tikz}
\usepackage{xcolor}

\makeatletter
\renewcommand{\eqref}[1]{\textup{\tagform@{\ref{#1}}}}
\makeatother

\definecolor{verifiedblue}{HTML}{2367A0}
\newif\ifshowverifiednumbers
\showverifiednumberstrue
\makeatletter
\newenvironment{verifiedequation}{%
  \begingroup
  \ifshowverifiednumbers
    \def\tagform@##1{\maketag@@@{\textcolor{verifiedblue}{(##1)}}}%
  \fi
  \begin{equation}%
}{%
  \end{equation}%
  \endgroup
}
\makeatother
\AtBeginDocument{%
  \setlength{\abovedisplayskip}{5pt plus 2pt minus 2pt}%
  \setlength{\belowdisplayskip}{5pt plus 2pt minus 2pt}%
  \setlength{\abovedisplayshortskip}{2pt plus 1pt minus 1pt}%
  \setlength{\belowdisplayshortskip}{3pt plus 1pt minus 1pt}%
  \setlength{\jot}{2pt}%
}
\nolinenumbers

\newcommand{\jj}{\mathrm{j}}
\newcommand{\CC}{\mathbb{C}}
\newcommand{\RR}{\mathbb{R}}

\makeatletter
\renewcommand{\p@subsection}{\thesection}
\renewcommand{\p@subsubsection}{\thesection}
\makeatother

\title{Minimum Effective-Susceptibility Bounds for Arbitrary Polarization Transformations in Metasurfaces}
\author{Ravi S. Hegde}
\affil{13/322 B, Department of Electrical Engineering, Indian Institute of Technology, Gandhinagar, India 382355, \texttt{hegder@iitgn.ac.in}}

\begin{abstract}
Metasurfaces are increasingly employed for sophisticated polarization transformations, including general unitary and nonunitary Jones operations. Despite substantial advances in analytical synthesis methods, inverse design of these transformations remains challenging. Here, we show that the particular mathematical structure of Jones polarization transformations makes optimal effective-susceptibility synthesis analytically tractable within the generalized sheet transition condition (GSTC) framework. For reciprocal, lossless dipolar metasurfaces, we establish exact feasibility and global minimum attainment, and derive sharp effective-susceptibility bounds for representative unitary and nonunitary transformations. We identify when magnetoelectric coupling lowers the minimum susceptibility, when it offers no resource advantage, and how the required material response depends on the prescribed polarization transformation. Applications to polarization rotations, tunable compensators, and polarization-state separation, including unambiguous discrimination, demonstrate sharp physical limits and material-resource trade-offs. These results provide geometry-independent theoretical benchmarks that can guide numerical inverse design and the selection of metasurface architectures.
\end{abstract}

\begin{document}
\maketitle

\section{Introduction}
\label{sec:introduction}

Metasurfaces have evolved from conventional polarization elements,
such as wave plates and polarization rotators, into versatile
platforms for polarization-dependent wavefront engineering.
Contemporary applications, including polarization-multiplexed
holography, vectorial beam shaping, and polarimetric imaging,
increasingly demand independent control over the amplitude,
phase, and polarization of multiple incident states
\cite{pfeiffer2014,rubin2021,shi2022,bao2022,arbabiPolarimetry2018}.
This progression has motivated a shift from designing individual
polarization conversions toward prescribing complete complex
Jones operators, encompassing both unitary and nonunitary
transformations. Full-parameter Jones-matrix synthesis provides
a general language for such functionality, including transformations
between arbitrary, potentially nonorthogonal polarization states
\cite{bao2022,chang2022universal,ji2024fullJones,yuan2024,guan2024}.
As the desired optical operations become more general, so too
does the need for systematic design methodologies capable of
translating arbitrary polarization specifications into physically
admissible metasurface responses.

The realization of increasingly general Jones operators has
been accompanied by a corresponding expansion in meta-atom
architectures and their accessible electromagnetic responses.
Early anisotropic single-layer elements, including dielectric
pillars and nanofins, primarily exploit birefringence to
implement wave-plate-like transformations
\cite{arbabi2015dielectric,rubin2021,shi2022}.
Greater control over polarization-dependent amplitude, phase,
and interchannel coupling has been achieved through coupled
resonators, rotated elements, bilayer structures, and stacked
metasurfaces \cite{bao2022,chang2022universal,yuan2024}.
More general bianisotropic architectures introduce
magnetoelectric coupling alongside electric and magnetic
responses, enlarging the space of attainable scattering
transformations
\cite{pfeiffer2014,pfeiffer2014asymmetric,epstein2016omega}.
These developments highlight an important distinction between
the electromagnetic degrees of freedom available to a metasurface
and those actually required for a prescribed optical operation.
In particular, the availability of additional constitutive
responses does not establish whether they are necessary for
realizing a transformation, or how substantially they reduce
its required effective susceptibility.

Numerical inverse design searches for meta-atom geometries
realizing prescribed optical responses, using adjoint,
evolutionary, and machine-learning approaches
\cite{molesky2018,inverseReview2022,hegdeMetagrating2022,
pandaHegde2022}. Such methods now address general
polarization transformations, including freeform
Jones-matrix meta-polarizers \cite{wang2024polarizers}.
Optimization also operates at the effective-sheet level:
Pearson and Hum optimized metasurface parameters under
far-field constraints \cite{pearsonHum2022}, while
Naseri \emph{et al.} coupled constrained bianisotropic
sheet optimization to multilayer unit-cell realization
\cite{naseri2022}. These approaches establish powerful
routes to functional designs. A distinct theoretical
question, however, is whether the prescribed optical
functionality admits a \emph{provable minimum} in its
constitutive requirements, independently of which
feasible geometry or scattering response is selected.

Generalized sheet transition conditions (GSTCs)
provide a well-established route from specified
electromagnetic fields and scattering responses to surface
susceptibilities. Foundational synthesis methods treat
prescribed incident, reflected, and transmitted fields,
including multiple independent transformations
\cite{achouri2015general,achouri2016illustrations}, while
bianisotropic scattering synthesis and polarization control
were developed by Pfeiffer and Grbic \cite{pfeiffer2014}.
Subsequent treatments systematized the constitutive freedoms
and the restrictions imposed by reciprocity, losslessness,
and symmetry \cite{achouri2018,budhuGrbic2021,
szymanski2021bianisotropic,achouri2021,cuestaAchouri2025}.

A recognized feature of GSTC synthesis is its
\emph{underdetermination}: even when electromagnetic
field transformations are prescribed, the boundary
equations may admit multiple constitutive realizations.
Foundational synthesis methods addressed this freedom
by restricting susceptibility components or imposing
additional independent transformations
\cite{achouri2015general,achouri2016illustrations,achouri2018}.
The freedom has also been exploited in design:
Achouri \emph{et al.} examined susceptibility-range
reduction through auxiliary phase and polarization
choices \cite{achouri2015range}, while subsequent
work optimized sheet parameters under far-field
constraints and connected them to physical unit cells
\cite{pearsonHum2022,naseri2022}.

Underdetermination thus provides a family of candidate
sheets for a prescribed functionality and raises the
question of \emph{which realization requires the least
constitutive resource}. Obtaining one feasible solution,
or improving a particular design numerically, does not
establish this minimum. A sharp bound requires optimization
over the full physically admissible family, together with
a proof that the resulting minimum is attained. In general,
however, the underdetermination of GSTC synthesis does not
itself make this optimization analytically tractable.

In this work, we find that prescribed Jones polarization
transformations possess a structure that makes this
optimization amenable to analytical treatment. The
specification of two input--output polarization states
and their transmitted efficiencies leaves transmission
phases and reflected fields undetermined, while the
two-dimensional Jones representation and the constraints
of reciprocity and losslessness strongly restrict the
admissible scattering responses. Exploiting this
structure within a homogeneous dipolar GSTC description,
we characterize the physically admissible bianisotropic
(BI) and non-bianisotropic (NB) sheet families and minimize
their effective susceptibility requirements without
imposing auxiliary phase or reflection prescriptions.

The analysis combines reciprocal two-port scattering
completions, inverse-Cayley susceptibility recovery,
and exact real-polynomial feasibility conditions.
These tools yield an exact semialgebraic characterization
of the minimum-susceptibility problem and rigorously
establish global minimum attainment for every feasible
prescribed transformation. For representative unitary
and nonunitary Jones transformations, we further obtain
sharp analytical susceptibility bounds, prove their
optimality, and construct attaining sheets. The results
expose class-dependent feasibility restrictions and
quantify when magnetoelectric coupling lowers the
constitutive resource requirement. In contrast to
established scattering-efficiency bounds
\cite{arbabi2017,abdelrahmanMonticone2023} and
related material-dependent $\mathbb{T}$-operator
limits \cite{molesky2022toperator}, our bounds quantify
the least effective-sheet susceptibility needed for
the prescribed polarization functionality. They provide
geometry-independent benchmarks for subsequent
susceptibility-constrained inverse design. Selected analytical statements are also formally
verified in Lean~4 under their stated assumptions
(blue equation numbers indicate formally verified relations);
the formal development is publicly available at
\url{https://github.com/rshegde/gstc-arbpol}.

The remainder of the paper is organized as follows. Section~2
introduces the polarization specification, susceptibility classes,
and feasibility formulation. Section~3 develops the analytical susceptibility
minima and optical applications, with supporting derivations
and optimality proofs in Supplement~1. Section~4
concludes the paper.

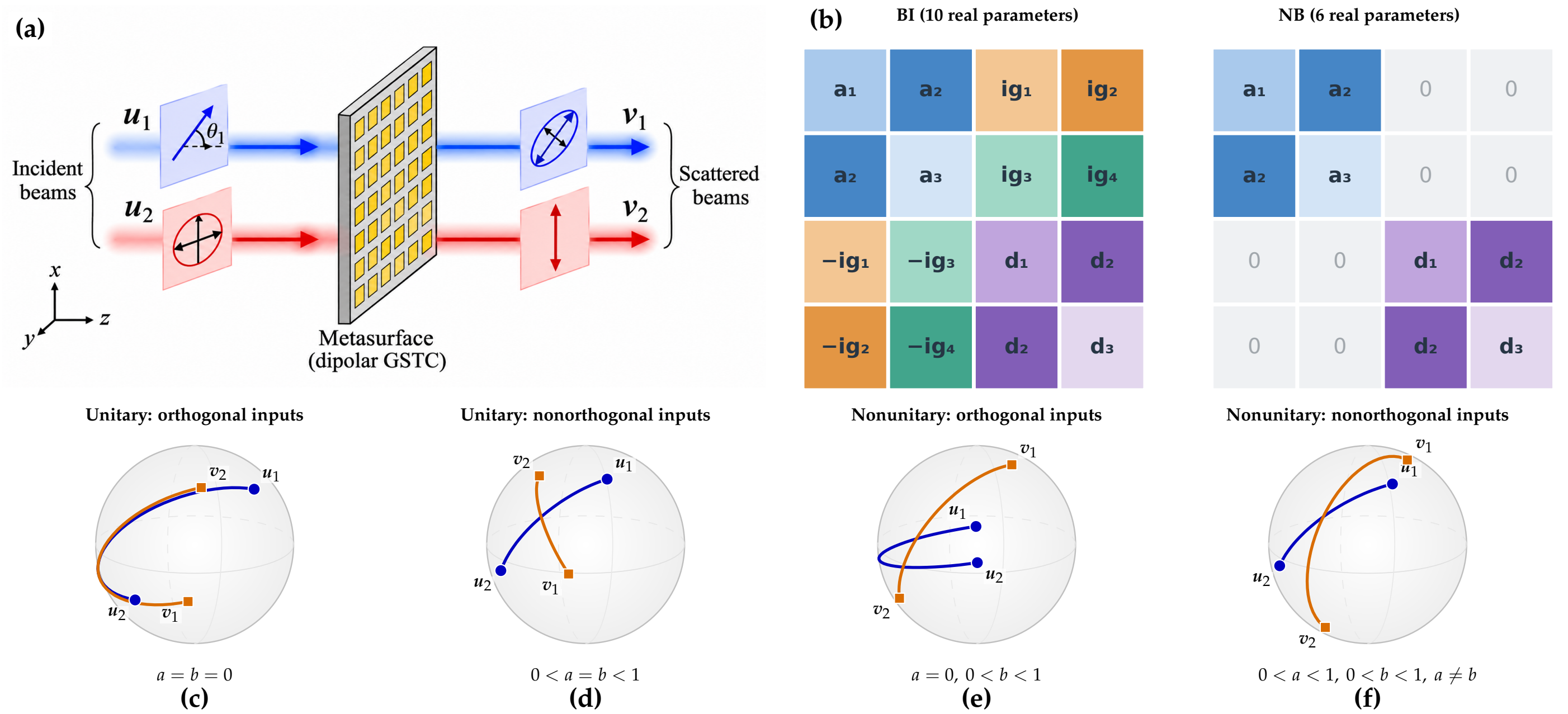
\begin{figure*}[!t]
\centering
\begin{minipage}[c]{0.49\textwidth}
  \centering
  \begin{tikzpicture}
    \node[inner sep=0,outer sep=0,anchor=south west] (panelA)
      at (0,0) {\includegraphics[width=\linewidth]{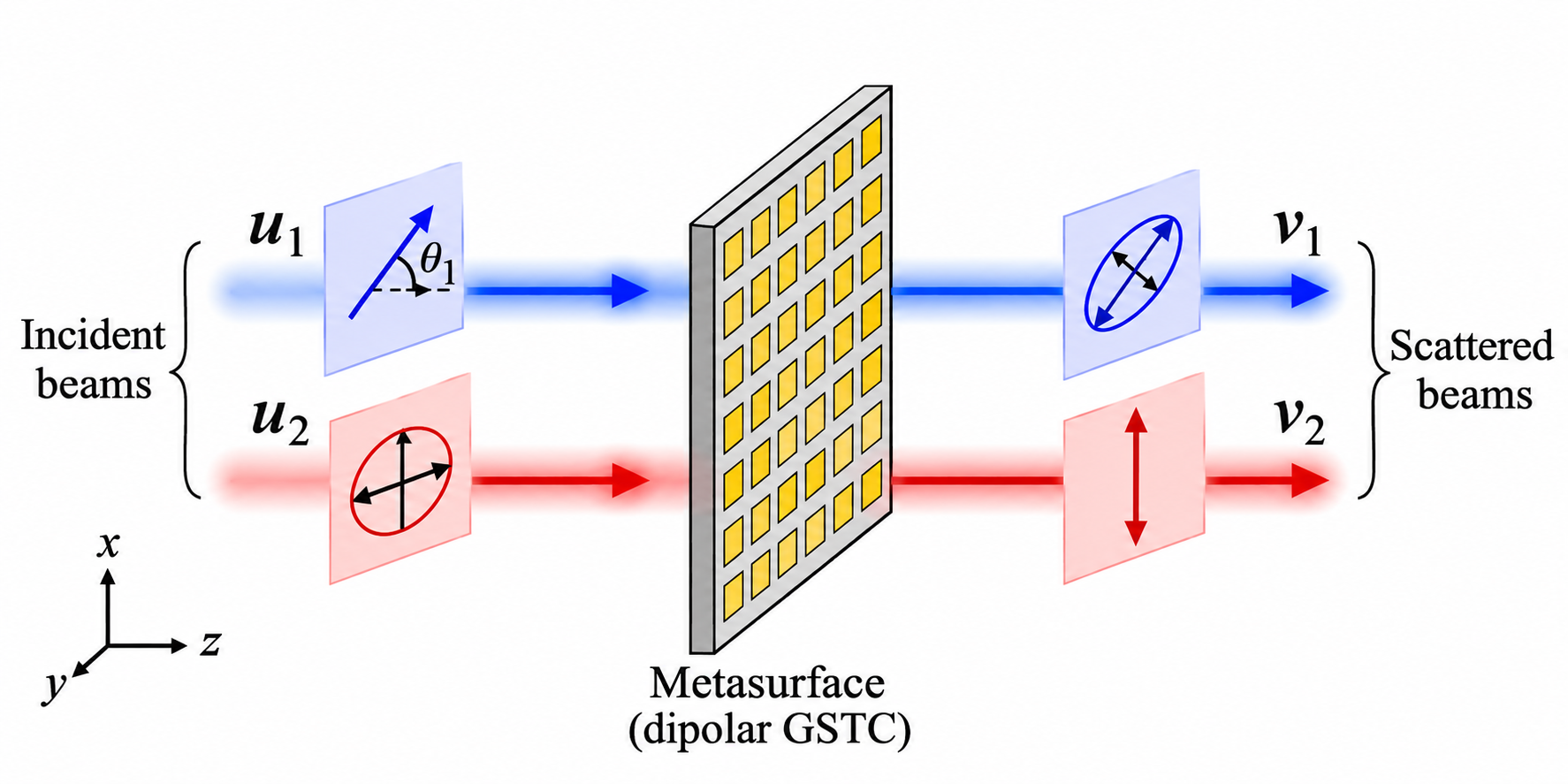}};
    \node[anchor=north west,font=\bfseries,fill=white,inner sep=2pt]
      at ([xshift=2pt,yshift=-2pt]panelA.north west) {(a)};
  \end{tikzpicture}
\end{minipage}\hfill
\begin{minipage}[c]{0.49\textwidth}
  \centering
  \begin{tikzpicture}
    \node[inner sep=0,outer sep=0,anchor=south west] (panelB)
      at (0,0) {\includegraphics[width=\linewidth]{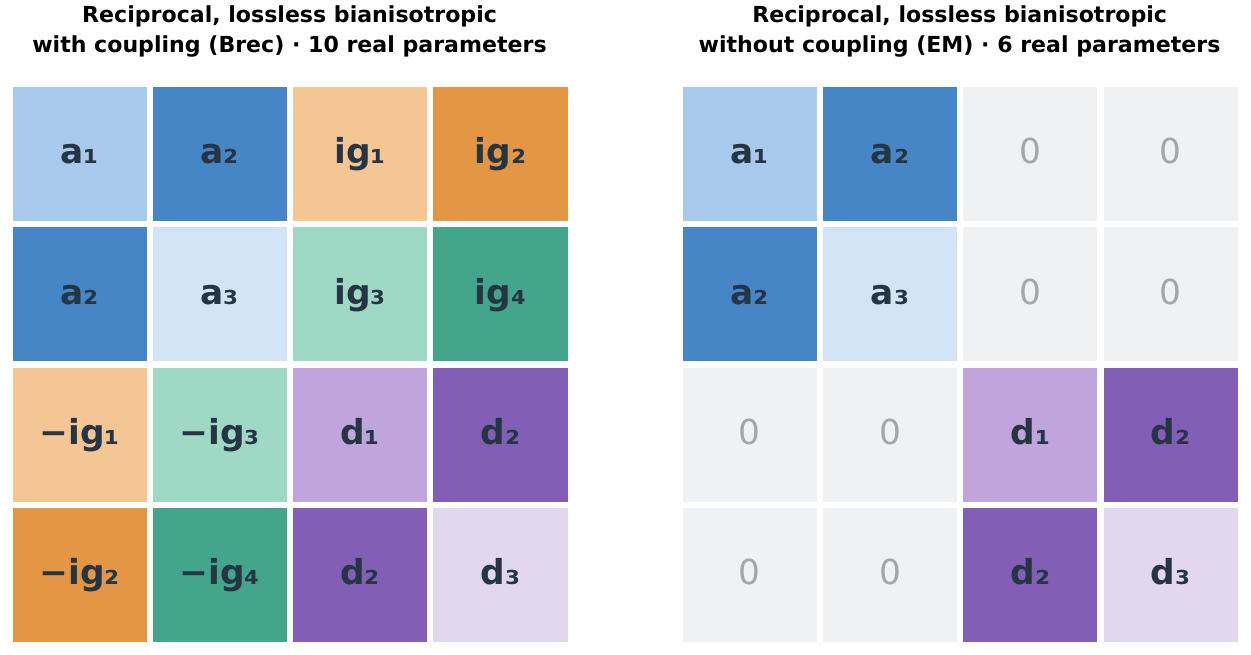}};
    \node[anchor=north west,fill=white,inner sep=0pt,
          minimum width=\linewidth,minimum height=.066\linewidth]
      at (panelB.north west) {};
    \node[anchor=north,font=\bfseries\scriptsize,inner sep=0pt]
      at ([xshift=.25\linewidth,yshift=-.014\linewidth]panelB.north west)
      {BI (10 real parameters)};
    \node[anchor=north,font=\bfseries\scriptsize,inner sep=0pt]
      at ([xshift=.75\linewidth,yshift=-.014\linewidth]panelB.north west)
      {NB (6 real parameters)};
    \node[anchor=north west,font=\bfseries,fill=white,inner sep=2pt]
      at ([xshift=2pt,yshift=-2pt]panelB.north west) {(b)};
  \end{tikzpicture}
\end{minipage}
\vspace{4pt}
\begin{minipage}[t]{0.247\textwidth}\centering
\begin{tikzpicture}[font=\scriptsize]
\node[font=\bfseries\scriptsize,anchor=north] at (0,1.73) {Unitary: orthogonal inputs};
\node[font=\bfseries\normalsize,anchor=north,inner sep=0pt] at (0,-1.68) {(c)};
\begin{scope}
\shade[inner color=white,outer color=gray!13] (0,0) circle (1.17);
\draw[gray!50,line width=.45pt] (0,0) circle (1.17);
\draw[gray!24,line width=.3pt] (-1.17,0) arc (180:360:1.17 and .34);
\draw[gray!20,line width=.3pt,dashed] (1.17,0) arc (0:180:1.17 and .34);
\draw[gray!18,line width=.3pt] (0,-1.17) arc (-90:90:.32 and 1.17);
\draw[gray!18,line width=.3pt,dashed] (0,1.17) arc (90:270:.32 and 1.17);
\draw[blue!75!black,line width=1pt] (0.704,0.655) -- (0.621,0.666) -- (0.532,0.672) -- (0.439,0.673) -- (0.342,0.668) -- (0.243,0.657) -- (0.142,0.642) -- (0.039,0.621) -- (-0.063,0.595) -- (-0.165,0.565) -- (-0.266,0.529) -- (-0.365,0.490) -- (-0.461,0.447) -- (-0.553,0.400) -- (-0.641,0.350) -- (-0.723,0.297) -- (-0.800,0.241) -- (-0.870,0.184) -- (-0.933,0.125) -- (-0.989,0.066) -- (-1.036,0.005) -- (-1.076,-0.055) -- (-1.107,-0.115) -- (-1.128,-0.174) -- (-1.141,-0.232) -- (-1.145,-0.287) -- (-1.139,-0.341) -- (-1.124,-0.391) -- (-1.100,-0.439) -- (-1.067,-0.483) -- (-1.026,-0.523) -- (-0.977,-0.559) -- (-0.919,-0.590) -- (-0.854,-0.617) -- (-0.782,-0.638) -- (-0.704,-0.655);
\draw[orange!85!black,line width=1pt] (-0.078,-0.674) -- (-0.180,-0.692) -- (-0.280,-0.705) -- (-0.379,-0.712) -- (-0.474,-0.713) -- (-0.565,-0.708) -- (-0.652,-0.698) -- (-0.734,-0.682) -- (-0.810,-0.661) -- (-0.879,-0.634) -- (-0.941,-0.603) -- (-0.995,-0.566) -- (-1.042,-0.525) -- (-1.080,-0.480) -- (-1.110,-0.430) -- (-1.130,-0.378) -- (-1.142,-0.322) -- (-1.144,-0.264) -- (-1.137,-0.203) -- (-1.121,-0.141) -- (-1.096,-0.078) -- (-1.062,-0.014) -- (-1.019,0.050) -- (-0.968,0.113) -- (-0.910,0.176) -- (-0.844,0.237) -- (-0.771,0.297) -- (-0.692,0.354) -- (-0.608,0.408) -- (-0.519,0.458) -- (-0.425,0.506) -- (-0.328,0.549) -- (-0.229,0.587) -- (-0.127,0.621) -- (-0.025,0.650) -- (0.078,0.674);
\filldraw[fill=blue!75!black,draw=white,line width=.4pt] (0.704,0.655) circle (.071);
\node[font=\scriptsize,anchor=south west,inner sep=.5pt,fill=white,fill opacity=.68,text opacity=1] at (0.774,0.720) {$\bm u_1$};
\filldraw[fill=blue!75!black,draw=white,line width=.4pt] (-0.704,-0.655) circle (.071);
\node[font=\scriptsize,anchor=north east,inner sep=.5pt,fill=white,fill opacity=.68,text opacity=1] at (-0.774,-0.720) {$\bm u_2$};
\filldraw[fill=orange!85!black,draw=white,line width=.4pt] (-0.138,-0.734) rectangle (-0.018,-0.614);
\node[font=\scriptsize,anchor=north east,inner sep=.5pt,fill=white,fill opacity=.68,text opacity=1] at (-0.148,-0.739) {$\bm v_1$};
\filldraw[fill=orange!85!black,draw=white,line width=.4pt] (0.018,0.614) rectangle (0.138,0.734);
\node[font=\scriptsize,anchor=south west,inner sep=.5pt,fill=white,fill opacity=.68,text opacity=1] at (0.148,0.739) {$\bm v_2$};
\end{scope}
\node[font=\scriptsize,anchor=north] at (0,-1.34) {$a=b=0$};
\end{tikzpicture}
\end{minipage}\hfill
\begin{minipage}[t]{0.247\textwidth}\centering
\begin{tikzpicture}[font=\scriptsize]
\node[font=\bfseries\scriptsize,anchor=north] at (0,1.73) {Unitary: nonorthogonal inputs};
\node[font=\bfseries\normalsize,anchor=north,inner sep=0pt] at (0,-1.68) {(d)};
\begin{scope}
\shade[inner color=white,outer color=gray!13] (0,0) circle (1.17);
\draw[gray!50,line width=.45pt] (0,0) circle (1.17);
\draw[gray!24,line width=.3pt] (-1.17,0) arc (180:360:1.17 and .34);
\draw[gray!20,line width=.3pt,dashed] (1.17,0) arc (0:180:1.17 and .34);
\draw[gray!18,line width=.3pt] (0,-1.17) arc (-90:90:.32 and 1.17);
\draw[gray!18,line width=.3pt,dashed] (0,1.17) arc (90:270:.32 and 1.17);
\draw[blue!75!black,line width=1pt] (0.254,0.774) -- (0.209,0.759) -- (0.164,0.743) -- (0.118,0.725) -- (0.072,0.706) -- (0.026,0.685) -- (-0.021,0.663) -- (-0.067,0.640) -- (-0.113,0.615) -- (-0.159,0.589) -- (-0.204,0.562) -- (-0.249,0.534) -- (-0.294,0.505) -- (-0.338,0.474) -- (-0.381,0.443) -- (-0.424,0.411) -- (-0.466,0.378) -- (-0.506,0.345) -- (-0.546,0.310) -- (-0.585,0.275) -- (-0.623,0.240) -- (-0.659,0.204) -- (-0.694,0.168) -- (-0.728,0.131) -- (-0.760,0.094) -- (-0.791,0.057) -- (-0.820,0.020) -- (-0.847,-0.018) -- (-0.873,-0.055) -- (-0.897,-0.092) -- (-0.919,-0.129) -- (-0.940,-0.166) -- (-0.958,-0.202) -- (-0.975,-0.238) -- (-0.990,-0.273) -- (-1.003,-0.308);
\draw[orange!85!black,line width=1pt] (-0.202,-0.348) -- (-0.226,-0.311) -- (-0.249,-0.273) -- (-0.272,-0.235) -- (-0.295,-0.197) -- (-0.317,-0.158) -- (-0.338,-0.119) -- (-0.359,-0.080) -- (-0.379,-0.040) -- (-0.398,-0.001) -- (-0.417,0.039) -- (-0.434,0.078) -- (-0.451,0.118) -- (-0.467,0.157) -- (-0.482,0.195) -- (-0.496,0.234) -- (-0.509,0.272) -- (-0.521,0.309) -- (-0.532,0.346) -- (-0.542,0.382) -- (-0.551,0.417) -- (-0.558,0.452) -- (-0.565,0.485) -- (-0.570,0.518) -- (-0.575,0.550) -- (-0.578,0.580) -- (-0.580,0.610) -- (-0.581,0.638) -- (-0.581,0.665) -- (-0.579,0.690) -- (-0.577,0.714) -- (-0.573,0.737) -- (-0.568,0.759) -- (-0.562,0.778) -- (-0.555,0.797) -- (-0.547,0.813);
\filldraw[fill=blue!75!black,draw=white,line width=.4pt] (0.254,0.774) circle (.071);
\node[font=\scriptsize,anchor=south west,inner sep=.5pt,fill=white,fill opacity=.68,text opacity=1] at (0.324,0.839) {$\bm u_1$};
\filldraw[fill=blue!75!black,draw=white,line width=.4pt] (-1.003,-0.308) circle (.071);
\node[font=\scriptsize,anchor=north east,inner sep=.5pt,fill=white,fill opacity=.68,text opacity=1] at (-1.073,-0.373) {$\bm u_2$};
\filldraw[fill=orange!85!black,draw=white,line width=.4pt] (-0.262,-0.408) rectangle (-0.142,-0.288);
\node[font=\scriptsize,anchor=north east,inner sep=.5pt,fill=white,fill opacity=.68,text opacity=1] at (-0.272,-0.413) {$\bm v_1$};
\filldraw[fill=orange!85!black,draw=white,line width=.4pt] (-0.607,0.753) rectangle (-0.487,0.873);
\node[font=\scriptsize,anchor=south east,inner sep=.5pt,fill=white,fill opacity=.68,text opacity=1] at (-0.617,0.878) {$\bm v_2$};
\end{scope}
\node[font=\scriptsize,anchor=north] at (0,-1.34) {$0<a=b<1$};
\end{tikzpicture}
\end{minipage}\hfill
\begin{minipage}[t]{0.247\textwidth}\centering
\begin{tikzpicture}[font=\scriptsize]
\node[font=\bfseries\scriptsize,anchor=north] at (0,1.73) {Nonunitary: orthogonal inputs};
\node[font=\bfseries\normalsize,anchor=north,inner sep=0pt] at (0,-1.68) {(e)};
\begin{scope}
\shade[inner color=white,outer color=gray!13] (0,0) circle (1.17);
\draw[gray!50,line width=.45pt] (0,0) circle (1.17);
\draw[gray!24,line width=.3pt] (-1.17,0) arc (180:360:1.17 and .34);
\draw[gray!20,line width=.3pt,dashed] (1.17,0) arc (0:180:1.17 and .34);
\draw[gray!18,line width=.3pt] (0,-1.17) arc (-90:90:.32 and 1.17);
\draw[gray!18,line width=.3pt,dashed] (0,1.17) arc (90:270:.32 and 1.17);
\draw[blue!75!black,line width=1pt] (-0.008,0.214) -- (-0.112,0.199) -- (-0.214,0.181) -- (-0.315,0.162) -- (-0.414,0.142) -- (-0.509,0.121) -- (-0.600,0.099) -- (-0.686,0.076) -- (-0.767,0.052) -- (-0.841,0.028) -- (-0.909,0.003) -- (-0.969,-0.021) -- (-1.022,-0.045) -- (-1.066,-0.069) -- (-1.102,-0.092) -- (-1.129,-0.115) -- (-1.147,-0.136) -- (-1.156,-0.157) -- (-1.155,-0.176) -- (-1.145,-0.194) -- (-1.126,-0.210) -- (-1.097,-0.225) -- (-1.060,-0.237) -- (-1.015,-0.248) -- (-0.961,-0.257) -- (-0.899,-0.264) -- (-0.830,-0.269) -- (-0.755,-0.271) -- (-0.673,-0.271) -- (-0.586,-0.269) -- (-0.495,-0.265) -- (-0.399,-0.259) -- (-0.300,-0.251) -- (-0.199,-0.241) -- (-0.096,-0.228) -- (0.008,-0.214);
\draw[orange!85!black,line width=1pt] (0.416,0.944) -- (0.365,0.934) -- (0.312,0.920) -- (0.258,0.903) -- (0.203,0.883) -- (0.148,0.859) -- (0.092,0.832) -- (0.035,0.801) -- (-0.022,0.768) -- (-0.078,0.732) -- (-0.135,0.693) -- (-0.191,0.651) -- (-0.246,0.607) -- (-0.300,0.560) -- (-0.353,0.512) -- (-0.404,0.461) -- (-0.455,0.409) -- (-0.503,0.355) -- (-0.549,0.300) -- (-0.594,0.243) -- (-0.636,0.186) -- (-0.676,0.128) -- (-0.713,0.070) -- (-0.747,0.011) -- (-0.779,-0.048) -- (-0.807,-0.106) -- (-0.833,-0.165) -- (-0.855,-0.222) -- (-0.874,-0.279) -- (-0.890,-0.335) -- (-0.903,-0.389) -- (-0.912,-0.442) -- (-0.917,-0.493) -- (-0.919,-0.542) -- (-0.918,-0.590) -- (-0.913,-0.635);
\filldraw[fill=blue!75!black,draw=white,line width=.4pt] (-0.008,0.214) circle (.071);
\node[font=\scriptsize,anchor=south east,inner sep=.5pt,fill=white,fill opacity=.68,text opacity=1] at (-0.078,0.279) {$\bm u_1$};
\filldraw[fill=blue!75!black,draw=white,line width=.4pt] (0.008,-0.214) circle (.071);
\node[font=\scriptsize,anchor=north west,inner sep=.5pt,fill=white,fill opacity=.68,text opacity=1] at (0.078,-0.279) {$\bm u_2$};
\filldraw[fill=orange!85!black,draw=white,line width=.4pt] (0.356,0.884) rectangle (0.476,1.004);
\node[font=\scriptsize,anchor=south west,inner sep=.5pt,fill=white,fill opacity=.68,text opacity=1] at (0.486,1.009) {$\bm v_1$};
\filldraw[fill=orange!85!black,draw=white,line width=.4pt] (-0.973,-0.695) rectangle (-0.853,-0.575);
\node[font=\scriptsize,anchor=north east,inner sep=.5pt,fill=white,fill opacity=.68,text opacity=1] at (-0.983,-0.700) {$\bm v_2$};
\end{scope}
\node[font=\scriptsize,anchor=north] at (0,-1.34) {$a=0,\ 0<b<1$};
\end{tikzpicture}
\end{minipage}\hfill
\begin{minipage}[t]{0.247\textwidth}\centering
\begin{tikzpicture}[font=\scriptsize]
\node[font=\bfseries\scriptsize,anchor=north] at (0,1.73) {Nonunitary: nonorthogonal inputs};
\node[font=\bfseries\normalsize,anchor=north,inner sep=0pt] at (0,-1.68) {(f)};
\begin{scope}
\shade[inner color=white,outer color=gray!13] (0,0) circle (1.17);
\draw[gray!50,line width=.45pt] (0,0) circle (1.17);
\draw[gray!24,line width=.3pt] (-1.17,0) arc (180:360:1.17 and .34);
\draw[gray!20,line width=.3pt,dashed] (1.17,0) arc (0:180:1.17 and .34);
\draw[gray!18,line width=.3pt] (0,-1.17) arc (-90:90:.32 and 1.17);
\draw[gray!18,line width=.3pt,dashed] (0,1.17) arc (90:270:.32 and 1.17);
\draw[blue!75!black,line width=1pt] (0.293,0.717) -- (0.246,0.704) -- (0.198,0.691) -- (0.151,0.676) -- (0.103,0.659) -- (0.055,0.642) -- (0.006,0.623) -- (-0.042,0.603) -- (-0.090,0.581) -- (-0.138,0.559) -- (-0.186,0.535) -- (-0.234,0.510) -- (-0.281,0.485) -- (-0.327,0.458) -- (-0.373,0.430) -- (-0.418,0.402) -- (-0.462,0.373) -- (-0.505,0.343) -- (-0.548,0.312) -- (-0.589,0.281) -- (-0.629,0.249) -- (-0.667,0.217) -- (-0.705,0.184) -- (-0.741,0.151) -- (-0.775,0.117) -- (-0.808,0.084) -- (-0.840,0.050) -- (-0.870,0.016) -- (-0.898,-0.018) -- (-0.924,-0.052) -- (-0.948,-0.086) -- (-0.971,-0.119) -- (-0.991,-0.153) -- (-1.010,-0.186) -- (-1.027,-0.219) -- (-1.041,-0.251);
\draw[orange!85!black,line width=1pt] (0.469,1.001) -- (0.419,1.023) -- (0.367,1.037) -- (0.311,1.043) -- (0.254,1.042) -- (0.194,1.032) -- (0.132,1.014) -- (0.070,0.988) -- (0.007,0.955) -- (-0.056,0.914) -- (-0.118,0.866) -- (-0.180,0.812) -- (-0.240,0.751) -- (-0.299,0.684) -- (-0.355,0.613) -- (-0.408,0.536) -- (-0.458,0.455) -- (-0.505,0.371) -- (-0.547,0.284) -- (-0.586,0.194) -- (-0.620,0.103) -- (-0.649,0.012) -- (-0.673,-0.080) -- (-0.692,-0.171) -- (-0.705,-0.261) -- (-0.713,-0.349) -- (-0.716,-0.434) -- (-0.713,-0.516) -- (-0.705,-0.593) -- (-0.691,-0.667) -- (-0.671,-0.735) -- (-0.647,-0.797) -- (-0.617,-0.853) -- (-0.583,-0.903) -- (-0.544,-0.945) -- (-0.501,-0.980);
\filldraw[fill=blue!75!black,draw=white,line width=.4pt] (0.293,0.717) circle (.071);
\node[font=\scriptsize,anchor=south west,inner sep=.5pt,fill=white,fill opacity=.68,text opacity=1] at (0.363,0.782) {$\bm u_1$};
\filldraw[fill=blue!75!black,draw=white,line width=.4pt] (-1.041,-0.251) circle (.071);
\node[font=\scriptsize,anchor=north east,inner sep=.5pt,fill=white,fill opacity=.68,text opacity=1] at (-1.111,-0.316) {$\bm u_2$};
\filldraw[fill=orange!85!black,draw=white,line width=.4pt] (0.409,0.941) rectangle (0.529,1.061);
\node[font=\scriptsize,anchor=south west,inner sep=.5pt,fill=white,fill opacity=.68,text opacity=1] at (0.539,1.066) {$\bm v_1$};
\filldraw[fill=orange!85!black,draw=white,line width=.4pt] (-0.561,-1.040) rectangle (-0.441,-0.920);
\node[font=\scriptsize,anchor=north east,inner sep=.5pt,fill=white,fill opacity=.68,text opacity=1] at (-0.571,-1.045) {$\bm v_2$};
\end{scope}
\node[font=\scriptsize,anchor=north] at (0,-1.34) {$0<a<1,\ 0<b<1,\ a\ne b$};
\end{tikzpicture}
\end{minipage}
\caption{Metasurface conversion, susceptibility classes, and Jones-operator geometry.
(a)~Metasurface schematic for the prescribed two-state transformations.
(b)~The BI and NB susceptibility matrices, with matching colors denoting repeated independent entries.
(c)--(f)~Four polarization-conversion cases represented by one Poincar\'e sphere each: unitary transformations of orthogonal (c) and nonorthogonal (d) input states; nonunitary transformations of orthogonal (e) and nonorthogonal (f) input states.
Blue circles and their geodesic arc denote the incident pair $\bm u_1,\bm u_2$; orange squares and their geodesic arc denote the target pair $\bm v_1,\bm v_2$.}
\label{fig:headline}
\end{figure*}

\section{Theoretical Framework}
\label{sec:theory}

As depicted in Fig.~\ref{fig:headline}(a), we prescribe
two transformations of normalized incident and target
Jones states $\bm u_j,\bm v_j\in\CC^2$, following
the Jones-matrix synthesis viewpoint \cite{bao2022}:
\begin{equation}
 \mathbf T\bm u_j=\alpha_j\bm v_j,\qquad
 \eta_j=|\alpha_j|^2,\qquad j=1,2.
 \label{eq:conversion}
\end{equation}
The target specifies the output polarization and
transmitted-power fraction $\eta_j$, but does not
prescribe the phases $\varphi_j=\arg\alpha_j$.
Such polarization-state requirements without an
additional phase-profile target have precedents
in metasurface design \cite{rubin2018polarimetry,yuan2024}. 
Only the incident states must be linearly independent;
the target states may coincide.
With $\mathbf U=[\bm u_1\ \bm u_2]$ invertible, all
transmission operators satisfying the two mappings
have the form
\begin{equation}
 \mathbf T=\mathbf V\operatorname{diag}(\alpha_1,\alpha_2)\mathbf U^{-1},
 \qquad \mathbf V=[\bm v_1\ \bm v_2].
 \label{eq:jones-operator-family}
\end{equation}
At fixed efficiencies, varying the unspecified
$\varphi_1,\varphi_2$ generally produces distinct Jones
operators for the \emph{same} polarization task.
Even for fixed $\mathbf T$, the reflected fields and
other two-port scattering blocks remain unspecified.
The prescribed polarization task therefore defines a
family of candidate scattering responses.

To classify these transformations, define the input
and output overlaps $d=\bm u_1^\dagger\bm u_2$
and $c=\bm v_1^\dagger\bm v_2$, with magnitudes
$a=|d|$ and $b=|c|$. Figures~\ref{fig:headline}(c)--(f) distinguish
orthogonal inputs ($a=0$) from nonorthogonal ones
($0<a<1$), and overlap-preserving ($a=b$) from
overlap-changing ($a\ne b$) transformations.
An overlap-preserving task admits a unitary Jones
representative; an overlap-changing task requires
nonunitary forward transmission when both
conversions have nonzero equal efficiency.
This does not imply absorption: the full lossless
two-port scattering operator remains unitary.

For passive transmission,
$\mathbf T^\dagger\mathbf T\preceq\mathbf I_2$.
The associated two-state Gram-matrix constraint is
standard \cite{chefles2002transformations,
cheflesJozsaWinter2004}. At common transmitted
efficiency $\eta_1=\eta_2=\eta$, its sharp ceiling is
\begin{equation}
 \eta_*(\mathcal T)=
 \begin{cases}
  (1-a)/(1-b),&a>b,\\
  1,&a=b,\\
  (1+a)/(1+b),&a<b.
 \end{cases}
 \label{eq:equal-efficiency-limit}
\end{equation}
This passive ceiling does not depend on a GSTC
material class. The complete two-port scattering
matrix must also satisfy reciprocity and losslessness,
which require symmetry and unitarity in the present
matched-port setting
\cite{achouri2021,szymanski2021bianisotropic}.

\subsection{GSTC model, susceptibility classes, and budgets}
\label{sec:gstc-model}

To realize the transformations depicted in
Fig.~\ref{fig:headline}(a), we model the metasurface
as a uniform, ultrathin, tangential dipolar sheet
at $z=0$, with identical free-space media on both
sides and excitation at normal incidence.
The standard generalized sheet transition conditions
(GSTCs) \cite{achouri2015general,pfeiffer2014,
budhuGrbic2021,achouri2021},
taken here as the starting electromagnetic model,
for $\exp(+\jj\omega t)$ time dependence, read
\begin{equation}
 \hat{\bm z}\times\Delta\bm H
   =\jj\omega\bm P_{\parallel},\qquad
 \hat{\bm z}\times\Delta\bm E
   =-\jj\omega\bm M_{\parallel},
 \label{eq:gstc}
\end{equation}
Here $\Delta$ denotes the jump across the sheet.
The corresponding surface constitutive relations are
\begin{equation}
 \begin{aligned}
 \bm P_\parallel
 &=\epsilon_0\bm\chi_{ee}\bm E_{\mathrm{av}}
   +c_0^{-1}\bm\chi_{em}\bm H_{\mathrm{av}},\\
 \bm M_\parallel
 &=c_0^{-1}\bm\chi_{me}\bm E_{\mathrm{av}}
   +\mu_0\bm\chi_{mm}\bm H_{\mathrm{av}},
 \end{aligned}
 \label{eq:gstc-constitutive}
\end{equation}
The subscript $\mathrm{av}$ denotes average
tangential fields, and $c_0$ is the free-space
light speed. We use the susceptibility convention of
Achouri and Martin \cite[Sec.~II]{achouri2021},
with magnetic surface polarization rescaled by
$\mu_0$. All susceptibilities have dimensions of length;
standard GSTC synthesis procedures are reviewed in
\cite{achouri2015general,achouri2018,budhuGrbic2021}.

We impose reciprocity and losslessness through
the following tangential BI parameterization
\cite{pfeiffer2014,achouri2021,szymanski2021bianisotropic,
achouriMartinPRB2021}:
\begin{verifiedequation}
 \begin{aligned}
 \bm\chi_{ee}&=\mathbf A=\mathbf A^{\mathsf T}\in\RR^{2\times2},&
 \bm\chi_{mm}&=\mathbf D=\mathbf D^{\mathsf T}\in\RR^{2\times2},\\
 \bm\chi_{em}&=\jj\mathbf G,&
 \bm\chi_{me}&=-\jj\mathbf G^{\mathsf T},
 \quad \mathbf G\in\RR^{2\times2}.
 \end{aligned}
 \label{eq:susceptibility-classes}
\end{verifiedequation}

As illustrated by the susceptibility matrices in
Fig.~\ref{fig:headline}(b), the bianisotropic (BI)
class has ten independent real parameters. The
non-bianisotropic (NB) subclass excludes
magnetoelectric coupling ($\mathbf G=\mathbf0$)
and retains six. Collecting the BI parameters gives
\begin{equation}
 \bm p=
 \bigl(A_{xx},A_{xy},A_{yy},
       D_{xx},D_{xy},D_{yy},
       G_{xx},G_{xy},G_{yx},G_{yy}\bigr)
 \in\RR^{10}.
 \label{eq:susceptibility-vector}
\end{equation}

To compare material requirements across classes,
we use the dimensionless, rotationally invariant
spectral susceptibility budget, with
$k_0=\omega/c_0=2\pi/\lambda_0$:
\begin{equation}
 B(\bm p)=k_0\max\{
       \|\mathbf A\|_2,\|\mathbf D\|_2,\|\mathbf G\|_2\}.
 \label{eq:sheet-budget}
\end{equation}
Here $\|\cdot\|_2$ is the matrix spectral norm,
i.e., the operator norm induced by the Euclidean
norm on tangential fields; it is not the Euclidean
norm of the ten-coordinate vector $\bm p$.
To distinguish direct electric--magnetic response
from magnetoelectric coupling within BI, define
\begin{equation}
 \begin{aligned}
 B_d(\bm p)&=k_0\max\{\|\mathbf A\|_2,
                              \|\mathbf D\|_2\},\\
 B_c(\bm p)&=k_0\|\mathbf G\|_2,\qquad
 B(\bm p)=\max\{B_d(\bm p),B_c(\bm p)\}.
 \end{aligned}
 \label{eq:two-budget-definition}
\end{equation}
Under a transverse rotation $\mathbf R$,
the susceptibility blocks transform as
$\mathbf M\mapsto\mathbf R^{\mathsf T}
\mathbf M\mathbf R$ for
$\mathbf M\in\{\mathbf A,\mathbf D,\mathbf G\}$;
their spectral norms are unchanged.
Moreover, rotating the incident and target Jones
states together with the sheet preserves both
the feasible family and its optimized minimum.
Thus $B$ provides a coordinate-independent measure
of the largest constitutive-block response. It is an
effective-sheet benchmark, not a direct bound on
fabricated meta-atom geometry. For fabrication-constrained
platforms, weighted electric, magnetic, and
magnetoelectric block costs can be considered;
the equal weighting here provides a matched-free-space
reference, not a universal measure of fabrication difficulty.

\subsection{Scattering--susceptibility relations}
\label{sec:six-equations}

The susceptibility classes restrict which scattering
responses can be realized by finite sheets. We leave
transmission phases and reflection unconstrained,
while enforcing reciprocity, losslessness, and the
GSTCs. For BI sheets, pole-free inverse-Cayley
recovery attains the passive efficiency ceiling,
even for coincident targets
(\href{supplement.pdf\#section.2}{Supplement~1, Sec.~S2}):
\begin{equation}
 \eta_{\max}^{\mathrm{BI}}(\mathcal T)
 =\eta_*(\mathcal T),\qquad
 \eta_{\max}^{\mathrm{NB}}(\mathcal T)
 \le\eta_*(\mathcal T).
 \label{eq:finite-bi-peak}
\end{equation}
For independent input and target Jones frames,
NB transmission symmetry imposes a further bound.
Define the cross-frame bilinear products
$h_{12}=\bm u_2^{\mathsf T}\bm v_1$ and
$h_{21}=\bm u_1^{\mathsf T}\bm v_2$.
At positive common efficiency, NB conversion is
excluded if $|h_{12}|\ne|h_{21}|$.
If both are nonzero and have equal magnitude,
$\Lambda=h_{12}/h_{21}$ fixes the relative
transmission phase. With
$\mathbf T_0=\mathbf V\operatorname{diag}(1,\Lambda)
\mathbf U^{-1}$, the NB peak is
\begin{verifiedequation}
 \eta_{\mathrm{NB}}^*=\|\mathbf T_0\|_2^{-2}.
 \label{eq:em-peak-transmission}
\end{verifiedequation}
If both products vanish, NB symmetry leaves the
relative phase free; the general feasibility
formulation \eqref{eq:six-equations}--\eqref{eq:minimum-budget}
still applies.

These efficiency limits characterize admissible
conversions, but not the susceptibility required
to realize them. The GSTC scattering--susceptibility
relations provide two complementary routes
\cite{pfeiffer2014,achouri2015general,achouri2021,szymanski2021bianisotropic}:
inverse recovery assigns a material budget to a
scattering response, whereas the forward relation
turns the Jones requirements into constraints on
the susceptibility parameters.
For a reciprocal, lossless two-port matrix $\mathbf S$,
let $\bm\Theta=\bigl(\begin{smallmatrix}\mathbf0&\mathbf I_2\\
\mathbf I_2&\mathbf0\end{smallmatrix}\bigr)$
exchange the ports. The inverse Cayley map is
\begin{verifiedequation}
 \mathbf K_{\mathbf S}=-2\jj
 (\bm\Theta\mathbf S-\mathbf I_4)
 (\bm\Theta\mathbf S+\mathbf I_4)^{-1}.
 \label{eq:main-inverse-cayley}
\end{verifiedequation}
The recovery map $\bm p=\mathcal P(\mathbf S)$ reads the ten real
susceptibility coordinates from the Hermitian $\mathbf K_{\mathbf S}$
when $\det(\bm\Theta\mathbf S+\mathbf I_4)\ne0$;
our normalization is given in \href{supplement.pdf\#section.1}{Supplement~1, Sec.~S1}.

For the forward direction, the GSTC scattering equations
give a rational transmission matrix
(\href{supplement.pdf\#section.1}{Supplement~1, Sec.~S1}):
\begin{verifiedequation}
 \mathbf T(\bm p)=\frac{\mathbf Q_T(\bm p)}{q_T(\bm p)},
 \qquad q_T(\bm p)\ne0,
 \label{eq:gstc-adjugate}
\end{verifiedequation}
Here $q_T$ and $\mathbf Q_T$ are polynomials in the
real susceptibility coordinates and $q_T\ne0$ for
every finite reciprocal, lossless sheet. Clearing this
nonvanishing denominator gives exact polynomial
conditions on the material parameters. With
$\bm w_j(\bm p)=\mathbf Q_T(\bm p)\bm u_j$,
output collinearity requires
\begin{verifiedequation}
 P_j(\bm p)=
 (v_j)_1(w_j)_2-(v_j)_2(w_j)_1=0.
 \label{eq:gstc-collinearity}
\end{verifiedequation}
The prescribed transmitted power requires
\begin{verifiedequation}
 R_j(\bm p;\eta_j)=
 \bm w_j^\dagger\bm w_j-\eta_j|q_T(\bm p)|^2=0,
 \qquad j=1,2.
 \label{eq:gstc-efficiency-polynomial}
\end{verifiedequation}
For each input, the complex collinearity condition
fixes the target polarization while the real power
condition fixes its efficiency. Together they give
six real-polynomial equations, without prescribing
the transmitted phases or imposing reflectionless
operation. Their necessity and sufficiency, including
at zero efficiency, are established in
\href{supplement.pdf\#section.1}{Supplement~1, Sec.~S1}:
\begin{verifiedequation}
 \begin{aligned}
 \operatorname{Re}P_1(\bm p)&=0,&
 \operatorname{Im}P_1(\bm p)&=0,\\
 \operatorname{Re}P_2(\bm p)&=0,&
 \operatorname{Im}P_2(\bm p)&=0,\\
 R_1(\bm p;\eta_1)&=0,&
 R_2(\bm p;\eta_2)&=0.
 \end{aligned}
 \label{eq:six-equations}
\end{verifiedequation}

\subsection{Minimum-susceptibility formulation}
\label{sec:minimum-formulation}

The six polynomial conditions of Sec.~\ref{sec:six-equations}
identify the sheets realizing a prescribed Jones
transformation at a specified efficiency. For the transformation
$\mathcal T=(\bm u_1,\bm u_2;\bm v_1,\bm v_2)$
at common efficiency $\eta_1=\eta_2=\eta$,
and susceptibility class
$\mathcal C\in\{\mathrm{BI},\mathrm{NB}\}$,
the feasible set is
\begin{equation}
 \begin{aligned}
 \mathcal F_{\mathcal C}(\mathcal T,\eta)
 =\{\bm p\in\mathcal C:\;&
 \operatorname{Re}P_j(\bm p)=0,\quad
 \operatorname{Im}P_j(\bm p)=0,\\
 &R_j(\bm p;\eta)=0,\quad j=1,2\}.
 \end{aligned}
 \label{eq:feasible-sheet-set}
\end{equation}
We minimize the susceptibility budget over this set:
\begin{equation}
 \begin{aligned}
 B_{\min}^{\mathcal C}(\mathcal T,\eta)
 &=\inf_{\bm p\in\mathcal C}\;
   \max\{B_d(\bm p),B_c(\bm p)\}\\
 \text{subject to}\quad
 &\operatorname{Re}P_j(\bm p)=
  \operatorname{Im}P_j(\bm p)=0,\\
 &R_j(\bm p;\eta)=0,\qquad j=1,2.
 \end{aligned}
 \label{eq:minimum-budget}
\end{equation}
Equivalently, this is $\inf_{\bm p\in\mathcal F_{\mathcal C}}B(\bm p)$;
we set $B_{\min}^{\mathcal C}=+\infty$ when the
feasible set is empty. Attainment is established
in Sec.~\ref{sec:general-minimum}.

To examine the trade-off between direct response and
magnetoelectric coupling, define the attainable
budget region
\begin{equation}
 \mathcal R_{\mathcal C}(\mathcal T,\eta)
 =\{(B_d(\bm p),B_c(\bm p)):
       \bm p\in\mathcal F_{\mathcal C}(\mathcal T,\eta)\}.
 \label{eq:budget-region}
\end{equation}
Its nondominated points form the Pareto frontier.
The separate minima
$B_{d,\min}^{\mathcal C}=\inf_{\bm p\in
\mathcal F_{\mathcal C}}B_d(\bm p)$ and
$B_{c,\min}^{\mathcal C}=\inf_{\bm p\in
\mathcal F_{\mathcal C}}B_c(\bm p)$ need not be attained by the same sheet, so they do
not generally determine $B_{\min}^{\mathcal C}$.
Instead, the objective minimizes $\max(B_d,B_c)$
over the attainable region. The six feasibility
equations, together with nine principal-minor
inequalities for the three spectral block-norm caps,
give a polynomial formulation used in the results below.

\subsection{General analytical results}
\label{sec:general-minimum}
We first establish attainment of the minimum,
then obtain the BI and NB minima at their respective
peak efficiencies through scattering completions,
without solving the full polynomial system
\eqref{eq:six-equations}.

\paragraph{Result 1: Attainment of the minimum.}
Every feasible transformation has a finite-sheet
realization that attains the least spectral budget:
\begin{verifiedequation}
 \begin{gathered}
 \{\beta\geq0:\mathcal H_{\mathcal C}
   (\beta;\mathcal T,\eta)\}
 =[B_{\min}^{\mathcal C}(\mathcal T,\eta),\infty),\\
 \exists\,\bm p_*\in\mathcal F_{\mathcal C}
  (\mathcal T,\eta):\quad
 B(\bm p_*)=B_{\min}^{\mathcal C}(\mathcal T,\eta).
 \end{gathered}
 \label{eq:budget-attainment}
\end{verifiedequation}
The compactness argument and algebraic budget-cap
test appear in \href{supplement.pdf\#section.2}{Supplement~1, Sec.~S2}.

At the peak feasible efficiency, saturation of the passivity bound generically makes the transmission power defect rank one, enabling the reduced scattering completion below. This simplification arises from passivity saturation rather than equal efficiency itself, and may also occur for unequal-efficiency tasks on the passive feasibility boundary.

\paragraph{Result 2: BI global minimum by scattering completion.}
For linearly independent input and target Jones pairs,
the transmission block at peak efficiency
$\eta_*(\mathcal T)$ has a rank-one power defect
(or zero defect at the unitary endpoint). Its
reciprocal, lossless scattering completions form a
complete family parameterized, in the generic
nonunitary case, by two phases: a common transmission
phase and a reflection phase. Inverse Cayley
recovery~\eqref{eq:main-inverse-cayley} gives
\begin{verifiedequation}
 B_{\min}^{\mathrm{BI}}(\mathcal T,\eta_*)
 =\min_{\mathbf S\in
   \mathscr S_{\mathrm{BI}}(\mathcal T,\eta_*)}
 B\bigl(\mathcal P(\mathbf S)\bigr).
 \label{eq:brec-general-budget}
\end{verifiedequation}
Here $\mathscr S_{\mathrm{BI}}$ is the complete
pole-free scattering family and $\mathcal P$ is
the inverse-Cayley susceptibility map. \href{supplement.pdf\#section.2}{Supplement~1, Sec.~S2} details the construction.
For dependent target pairs, the BI efficiency ceiling
remains attainable, but the nonsingular scattering
parameterization used for this minimum-budget formula
does not apply; the general formulation
\eqref{eq:six-equations}--\eqref{eq:minimum-budget}
remains valid.

\paragraph{Result 3: NB global minimum by parity completion.}
For nonzero cross-frame bilinear products, let
$\mathbf T_*=\sqrt{\eta_{\mathrm{NB}}^*}\,\mathbf T_0$,
using the NB peak established in
Sec.~\ref{sec:six-equations}.
Minimizing over the complete symmetric-unitary parity
completions $\mathscr P_{\mathrm{NB}}(\mathbf T_*)$,
followed by inverse Cayley recovery, gives
\begin{verifiedequation}
 \begin{aligned}
 B_{\min}^{\mathrm{NB}}
  (\mathcal T,\eta_{\mathrm{NB}}^*)
 =k_0\!
 \min_{\substack{(\mathbf X,\mathbf Y)\in
                  \mathscr P_{\mathrm{NB}}(\mathbf T_*)\\
                 |z|=1,\ \mathrm{pole\!-\!free}}}
 \max\!\bigl\{
       \|\mathbf A(z,\mathbf X)\|_2,
       \|\mathbf D(z,\mathbf Y)\|_2\bigr\}.
 \end{aligned}
 \label{eq:em-general-budget}
\end{verifiedequation}
\href{supplement.pdf\#section.2}{Supplement~1, Sec.~S2} gives the parity
construction and recovery maps. The general
formulation \eqref{eq:six-equations}--\eqref{eq:minimum-budget} also covers vanishing cross-frame bilinear products
and other feasible efficiencies.

\section{Results and Discussion}
\label{sec:results}
We apply the minimum-budget results of
Sec.~\ref{sec:general-minimum} first to unitary
polarization transformations and then to nonunitary
tasks and their efficiency--susceptibility trade-offs.
Because the BI and NB efficiency ceilings need not
coincide, we compare their susceptibility budgets
at a common efficiency only when that efficiency
is attainable in both classes.

\subsection{Unitary polarization transformations}
\label{sec:unitary}

At unit efficiency, Results~2 and~3 of
Sec.~\ref{sec:general-minimum} simplify the
minimum-susceptibility problem for unitary Jones
transformations. The key distinction is how much
transmission-phase freedom the prescribed states leave.
The final class budgets follow from globally complete
scattering constructions; \href{supplement.pdf\#section.3}{Supplement~1, Sec.~S3}
establishes the phase centering and optimality.

For orthonormal input and output bases,
$\mathbf T=\mathbf V\operatorname{diag}
(e^{\jj\phi_1},e^{\jj\phi_2})\mathbf U^\dagger$.
The relative transmission phase remains free.
The determinant-centered Jones matrix can be written
$\mathbf M(t)=q(t)\mathbf I_2+
\jj\bm r(t)\cdot\bm\sigma$.
Here $t$ parameterizes that free phase and the four
real coordinates $q(t),r_x(t),r_y(t),r_z(t)$
are fixed trigonometric functions of $t$ once
the prescribed polarization states are chosen.
Minimizing over the common phase is analytic,
leaving the direct electric--magnetic and
magnetoelectric costs of the same centered sheet:

\begin{verifiedequation}
 \begin{aligned}
 B_d(t)&=\frac{2w\sqrt{r_x(t)^2+r_z(t)^2}}{1+|q(t)|},\\
 B_c(t)&=\frac{2v|r_y(t)|}{1+|q(t)|},\\
 B(t)&=\max\{B_d(t),B_c(t)\},
 \end{aligned}
 \label{eq:orth-budget}
\end{verifiedequation}
Here $w=v=1$ in free space. Neither component
is an independently optimized budget: for each
relative phase the two costs belong to the same
attaining sheet. BI may choose any phase and
minimizes their upper envelope, whereas NB must
choose a phase with vanishing coupling,
$r_y(t)=0$. The resulting globally sharp minima are

\begin{verifiedequation}
 B_{\min}^{\mathrm{BI}}(\mathcal T)
   =\min_t B(t),
 \label{eq:orth-brec-min}
\end{verifiedequation}
\begin{verifiedequation}
 B_{\min}^{\mathrm{NB}}(\mathcal T)
   =\min_{t:\,r_y(t)=0} B(t).
 \label{eq:orth-em-min}
\end{verifiedequation}
This is the central simplification: an optimization
over all admissible sheet susceptibilities has
become a one-variable phase minimization.
We measure the BI advantage by
\begin{equation}
 \mathcal S(\mathcal T)=
 1-\frac{B_{\min}^{\mathrm{BI}}(\mathcal T)}
         {B_{\min}^{\mathrm{NB}}(\mathcal T)},
 \label{eq:orth-saving}
\end{equation}
defined for $B_{\min}^{\mathrm{NB}}>0$.
Class inclusion gives $0\leq\mathcal S\leq1$;
the absolute saving $\Delta B=
B_{\min}^{\mathrm{NB}}-B_{\min}^{\mathrm{BI}}$
obeys the sharp bound
\begin{equation}
 0\leq\Delta B\leq 2w .
 \label{eq:orth-absbound}
\end{equation}
The phase-global proof and a family approaching this bound
are given in \href{supplement.pdf\#section.3}{Supplement~1, Sec.~S3}.

In contrast, for independent nonorthogonal input states with
$0<a=b<1$, equal-efficiency unitarity fixes the
relative output phase through the complex overlaps
$d=\bm u_1^\dagger\bm u_2$ and
$c=\bm v_1^\dagger\bm v_2$.
Define the transmission representative
$\mathbf X=\mathbf V\operatorname{diag}(1,e^{\jj\psi})
\mathbf U^{-1}$. The overlap-preserving phase and
resulting unitarity satisfy
\begin{verifiedequation}
 e^{\jj\psi}c=d,\qquad
 \mathbf X^\dagger\mathbf X=\mathbf I_2.
 \label{eq:unitary-nonorth-phase}
\end{verifiedequation}
The relative-phase optimization has disappeared; only a common transmission phase remains, and its optimum is analytic.
For the determinant-centered representative
$\mathbf X_c=q\mathbf I_2+\jj\bm r\cdot\bm\sigma$,
the sharp free-space BI minimum is
(\href{supplement.pdf\#section.3}{Supplement~1, Sec.~S3})
\begin{equation}
 B_{\min}^{\mathrm{BI}}(\mathcal T,1)
  =\frac{2\max\{\sqrt{r_x^2+r_z^2},|r_y|\}}{1+|q|}.
 \label{eq:nonorth-unitary-min}
\end{equation}
The NB class additionally requires the reciprocal
cross-frame condition derived in
\href{supplement.pdf\#section.2}{Supplement~1, Sec.~S2},
equivalent here to $\mathbf X^{\mathsf T}=\mathbf X$
or $r_y=0$. Because the relative phase is fixed,
this condition cannot be restored by changing the
common phase. Consequently,
\begin{equation}
 B_{\min}^{\mathrm{NB}}(\mathcal T,1)=
 \begin{cases}
 B_{\min}^{\mathrm{BI}}(\mathcal T,1),&
 \mathbf X^{\mathsf T}=\mathbf X,\\
 +\infty,&\mathbf X^{\mathsf T}\ne\mathbf X .
 \end{cases}
 \label{eq:nonorth-unitary-class-comparison}
\end{equation}
Thus NB has no finite budget for a nonsymmetric
representative, whereas the BI minimum remains finite.
For every unitary Jones operation the optimized BI scalar
budget obeys the sharp universal bound
$B_{\min}^{\mathrm{BI}}\leq2$ in free space
(and likewise for feasible NB operations).
This dimensionless ceiling on the minimum
block-spectral susceptibility budget follows from
$q^2+|\bm r|^2=1$; it is not a material
susceptibility limit.

The following applications show the relative-phase
trade-off in orthogonal linear-to-elliptical conversion,
and the explicit budget's implications for a
continuously tunable compensator.

\subsubsection{Linear-to-elliptical conversion}
\label{sec:linear-elliptical}

Wave-plate-like transformations are central to
metasurface polarization control, from linear
rotations to conversion between linear and
circular states. Dielectric metasurface wave
plates have demonstrated efficient polarization
manipulation
\cite{kruk2016broadband,he2022metawaveplate},
while linearly birefringent meta-atoms have
provided independent control of orthogonal
linear, circular, and elliptical channels
\cite{mueller2017orthogonal}. These devices
motivate a precise material-cost comparison:
for a prescribed unit-efficiency transformation,
how much direct electric--magnetic response is
needed, and when can magnetoelectric coupling
reduce that requirement?

We take two orthogonal linear inputs and
prescribe an arbitrary orthogonal output pair.
With $\mathbf U=\mathbf I_2$, the output
polarizations span the Poincar\'e sphere:
\begin{equation}
 \begin{gathered}
 \mathbf V=[\,\bm e(\chi,\phi)\
                 \bm e(\chi+\pi,\phi)\,],\\
 \bm e(\chi,\phi)=
 \bigl(\cos(\chi/2),\,
       e^{\jj\phi}\sin(\chi/2)\bigr)^{\mathsf T}.
 \end{gathered}
 \label{eq:orth-elliptical}
\end{equation}
The $\phi=0$ slice describes linear-output
transformations, while
$(\chi,\phi)=(\pi/2,\pm\pi/2)$ gives
circular outputs. Both BI and NB are optimized
over all admissible relative transmission phases
for the same prescribed input--output pairs.

Consider first a $30^\circ$ linear rotation. On
$0\le t\le\pi/2$, Eq.~\eqref{eq:orth-budget}
reduces to
$B_d(t)=2\sin t/(1+\cos30^\circ\cos t)$
and $B_c(t)=\cos t/(1+\cos30^\circ\cos t)$.
Changing the otherwise unspecified relative
phase trades direct susceptibility against
magnetoelectric coupling. The BI optimum
balances the two costs at their intersection,
$t=\tan^{-1}(1/2)$, where
$B_{\min}^{\mathrm{BI}}\simeq0.5040$.
NB cannot use that phase because it requires
nonzero coupling; its symmetric transmission
condition selects $t=\pi/2$, giving
$B_{\min}^{\mathrm{NB}}=2$. Thus the BI sheet
requires $74.8\%$ less minimum effective
susceptibility for exactly the same rotation,
rather than merely achieving a different
phase response or conversion efficiency.

\begin{figure*}[!t]
\centering
\begin{tikzpicture}
\node[inner sep=0,outer sep=0,anchor=south west] (overview)
  at (0,0)
  {\includegraphics[width=\textwidth]{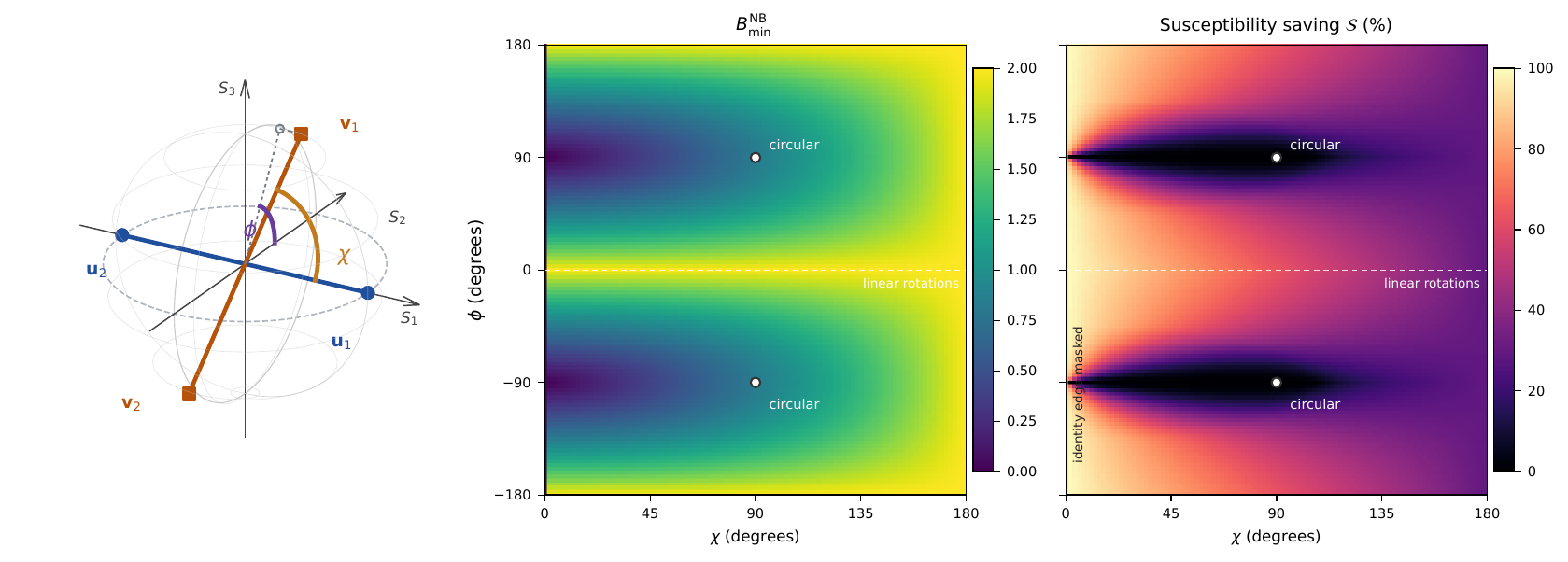}};
\node[anchor=north west,font=\bfseries] at
  ([xshift=0.01\textwidth,yshift=-0.02cm]overview.north west) {(a)};
\node[anchor=north west,font=\bfseries] at
  ([xshift=0.345\textwidth,yshift=-0.02cm]overview.north west) {(b)};
\node[anchor=north west,font=\bfseries] at
  ([xshift=0.680\textwidth,yshift=-0.02cm]overview.north west) {(c)};
\end{tikzpicture}
\caption{Minimum rotationally invariant spectral
susceptibility budget for unit-efficiency
linear-to-elliptical conversion between orthogonal
polarization bases. (a)~Poincar\'e-sphere geometry;
(b)~minimum NB spectral budget;
(c)~fractional susceptibility saving $\mathcal S$
obtained with BI sheets. White circles identify
circular-output states, where the BI and NB
spectral minima coincide; the dashed line
marks linear-rotation transformations. The identity
edge is masked in panel (c), where the fractional
saving is undefined because both minima vanish.
The plotted budgets are globally minimized over
admissible phases and scattering completions under
the spectral objective; the centered realizations
attain these minima (\href{supplement.pdf\#section.3}{Supplement~1, Sec.~S3}).}
\label{fig:linear}
\end{figure*}
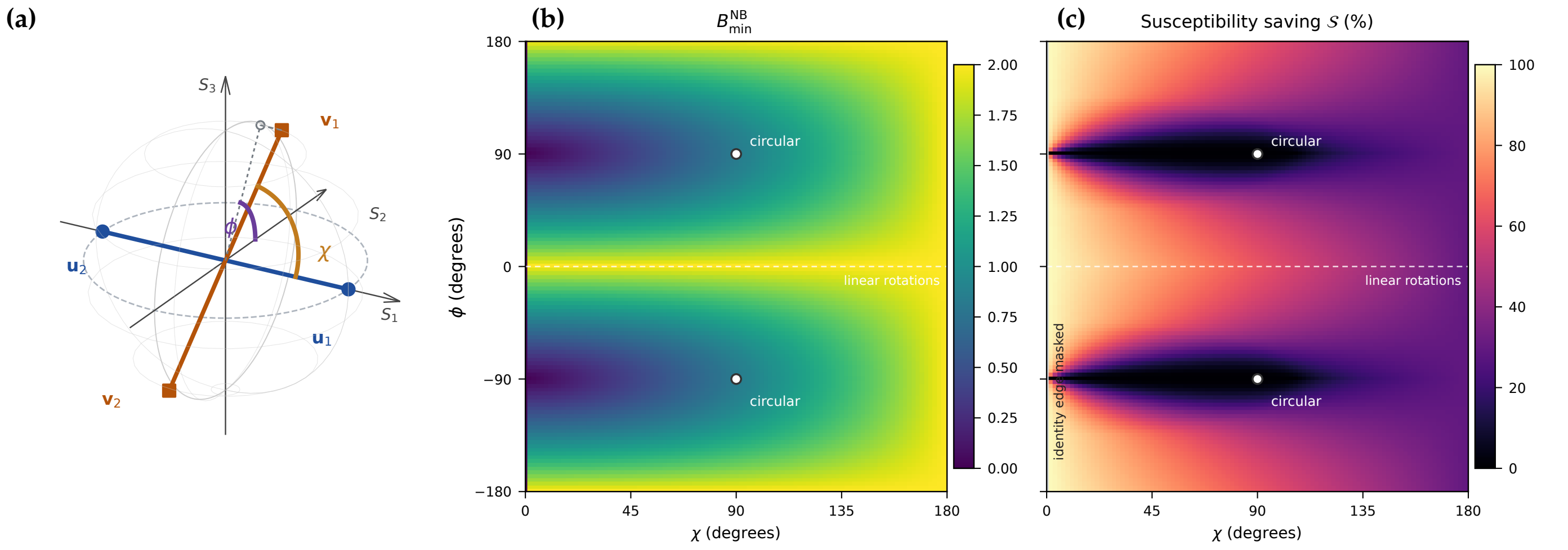

Repeating the global phase minimization at each
$(\chi,\phi)$ gives Fig.~\ref{fig:linear}:
the NB minimum susceptibility and the
fractional reduction attainable in BI.
The landscape shows that additional constitutive
freedom does not provide a uniform advantage.
For linear-to-circular conversion, both classes
have the same sharp minimum,
$B_{\min}^{\mathrm{BI}}=B_{\min}^{\mathrm{NB}}
=2(\sqrt2-1)$. By contrast, the linear-rotation
slice exhibits substantial savings, culminating
in a singular limit of the \emph{optimized NB
budget} near the identity.

To understand its origin, consider a linear
rotation through $\theta$ with an unspecified
relative transmission phase $2t$. Up to an
irrelevant common phase, the Jones matrix is
$\mathbf T=R(\theta)
\operatorname{diag}(e^{\jj t},e^{-\jj t})$.
For an NB sheet, the forward transmission
must be symmetric,
$\mathbf T^{\mathsf T}=\mathbf T$.
At any nonzero rotation this requires
$\cos t=0$: the two linear input channels
acquire a relative phase of $\pi$, even when
the prescribed rotation angle is arbitrarily
small. In this sense the NB representative
is half-wave-like, with a finite retardance
whose principal axes change with $\theta$.
Its minimum spectral budget therefore remains
$B_{\min}^{\mathrm{NB}}=2$ as
$\theta\to0$ through nonzero rotations.
At the exact identity, however, the symmetry
constraint no longer fixes the relative phase;
the transparent sheet becomes admissible
and the minimum drops to zero.

BI sheets avoid this constraint because
magnetoelectric coupling permits a nonsymmetric
forward transmission. Their phase-optimal
representatives can approach the identity
transmission and zero susceptibility as
$\theta\to0$. Thus the discontinuity is
not a singularity of the electromagnetic
fields, but a change in the admissible
transmission-phase choices. The sharp limits
are established in
\href{supplement.pdf\#section.3}{Supplement~1, Sec.~S3}.

This distinction is relevant to a tunable
linear rotator operating around zero rotation.
An NB sheet can in principle tune continuously
through zero by remaining on a finite-response,
half-wave-like branch, including at the
identity. What it cannot do is follow the
pointwise minimum \emph{at} the identity while
also approaching that transparent solution
continuously from nonzero rotations.
A BI design, in contrast, has access to a
low-susceptibility branch approaching transparency.
This is a sheet-level material-resource advantage,
not by itself a claim about tuning voltage
or an achievable bias trajectory.

\subsubsection{Tunable transformations: susceptibility capacity}
\label{sec:unitary-rotations}

In polarization-encoded quantum communication, protocols such as
B92 employ nonorthogonal states \cite{bennett1992b92}.
Fiber birefringence can apply a time-varying unitary rotation
that preserves their overlap but changes their orientations,
motivating adaptive polarization compensation
\cite{chen2024realtimepol}.
As an application of the explicit nonorthogonal unitary budget above,
we consider a reciprocal, lossless BI metasurface compensator
and determine the susceptibility capacity needed for a
continuously tunable family of unitary rotations,
without assuming a particular physical control mechanism.

For a Poincar\'e-sphere rotation axis
$\hat{\bm n}=(n_x,n_y,n_z)$, $|\hat{\bm n}|=1$,
we prescribe
\begin{equation}
 \mathbf U(\hat{\bm n},\Theta)
 =\cos(\Theta/2)\mathbf I_2
 -\jj\sin(\Theta/2)\hat{\bm n}\cdot\bm\sigma,
 \quad 0\leq\Theta\leq\pi .
 \label{eq:unitary-rotation}
\end{equation}
The Pauli coordinates $(n_x,n_y,n_z)$ correspond
to the $(S_2,S_3,S_1)$ Stokes directions.
Thus $\hat{\bm n}$ is a polarization-space rotation
axis, not the normal to the physical sheet.
For each selected axis, the continuous family begins
at transparent transmission for $\Theta=0$ and reaches
half-wave retardance at $\Theta=\pi$. Varying the
axis as well as the angle spans the ideal lossless
polarization-compensation operations, up to a
physically irrelevant common Jones phase.

Specializing the globally minimized unitary budget
to the rotation-axis coordinates gives the
following sharp material requirements:
\begin{equation}
 \begin{aligned}
 B_{d,\mathrm{cent}}(\hat{\bm n},\Theta)
   &=f(\Theta)\sqrt{n_x^2+n_z^2},\\
 B_{c,\mathrm{cent}}(\hat{\bm n},\Theta)
   &=f(\Theta)|n_y|,\\
 B_{\min}^{\mathrm{BI}}(\hat{\bm n},\Theta)
   &=f(\Theta)\max\{\sqrt{1-n_y^2},|n_y|\},\\
 f(\Theta)&=2\tan(\Theta/4).
 \end{aligned}
 \label{eq:rotation-budget}
\end{equation}
The two block costs describe the same
globally optimal centered realization, rather
than independently optimized resources.
The phase-centering proof is in
\href{supplement.pdf\#section.3}{Supplement~1, Sec.~S3}.

At a fixed axis, this minimum budget increases
monotonically with the rotation angle $\Theta$.
Thus the peak susceptibility capacity along the
tuning trajectory occurs at the half-wave endpoint:

\begin{equation}
 \begin{aligned}
 B_{\mathrm{cap}}^*(\hat{\bm n})
 &=\sup_{0\leq\Theta\leq\pi}
      B_{\min}^{\mathrm{BI}}(\hat{\bm n},\Theta)\\
 &=B_{\min}^{\mathrm{BI}}(\hat{\bm n},\pi)
  =2\max\{\sqrt{1-n_y^2},|n_y|\},\\
 D_{\mathrm{cap,cent}}
 &=2\sqrt{1-n_y^2},\qquad
 C_{\mathrm{cap,cent}}=2|n_y|.
 \end{aligned}
 \label{eq:unitary-tuning-capacity}
\end{equation}
At quarter-wave retardance the centered minimum
is $(\sqrt2-1)g(\hat{\bm n})\,2$;
at half-wave retardance it is $2g(\hat{\bm n})$.
The half-wave capacity ranges from $\sqrt2$
(for $|n_y|=1/\sqrt2$) to $2$
(for $|n_y|=0$ or $1$). A linear-eigenaxis
retarder ($n_y=0$) uses only direct electric
and magnetic response in this realization;
a circular-eigenaxis rotator ($|n_y|=1$)
uses only magnetoelectric coupling; mixed
eigenaxes require both. These are distinct
optical operations, not interchangeable
realizations of one Jones operator.

For a compensator that can select any axis and any
retardance, the corresponding ideal pointwise
capacity has the universal envelope
\begin{equation}
 B_{\mathrm{cap,all}}^*
 =\sup_{|\hat{\bm n}|=1}\sup_{0\leq\Theta\leq\pi}
 B_{\min}^{\mathrm{BI}}(\hat{\bm n},\Theta)=2.
 \label{eq:rotation-universal-capacity}
\end{equation}
This is the minimum \emph{scalar peak} budget
for a reconfigurable collection of ideal sheets:
it should not be confused with the electronic
control range or with a claim about a particular
physical tuning mechanism. Within the same
centered construction, the direct and
magnetoelectric block peak capacities are
$D_{\mathrm{cap,cent}}=2\sqrt{1-n_y^2}$ and
$C_{\mathrm{cap,cent}}=2|n_y|$, respectively.
Consequently both block types must be available
to span arbitrary axes, with separate maxima of
$2$ over the complete axis sphere. Their maxima
occur at different axes, so these two capacities
must not be added to form the scalar optimum.

\begin{figure*}[!t]
\centering
\begin{minipage}[c]{0.48\textwidth}
\centering
\begin{tikzpicture}[x=0.93cm,y=0.93cm]
\foreach \ix in {0,...,47}{
 \foreach \iy in {0,...,23}{
  \pgfmathsetmacro{\thdeg}{(\ix+0.5)*180/48}
  \pgfmathsetmacro{\qy}{(\iy+0.5)/24}
  \pgfmathsetmacro{\gg}{max(sqrt(max(0,1-\qy*\qy)),\qy)}
  \pgfmathsetmacro{\bb}{2*tan(\thdeg/4)*\gg}
  \pgfmathsetmacro{\pct}{50*\bb}
  \pgfmathsetmacro{\xx}{\ix/8}
  \pgfmathsetmacro{\yy}{\iy/8}
  \fill[blue!\pct!white] (\xx,\yy) rectangle ++(0.125,0.125);
 }}
\begin{scope}
 \clip (0,0) rectangle (6,3);
 \foreach \lvl in {0.5,1,1.5}{
  \draw[white,thick,domain=0:1,samples=100,smooth,variable=\qq]
   plot ({4*atan(\lvl/(2*max(sqrt(max(0,1-\qq*\qq)),\qq)))/30},{3*\qq});
 }
\end{scope}
\draw (0,0) rectangle (6,3);
\foreach \x/\t in {0/0,3/{\pi/2},6/{\pi}}
 \node[font=\scriptsize,below] at (\x,0) {$\t$};
\foreach \y/\t in {0/0,1.5/{0.5},3/1}
 \node[font=\scriptsize,left] at (0,\y) {$\t$};
\node[font=\small] at (3,-0.55) {Retardance $\Theta$};
\node[font=\small,rotate=90] at (-0.58,1.5) {$|n_y|$};
\node[font=\small] at (3,3.45) {(a) $B_{\min}^{\mathrm{BI}}$: 0 (white) to 2 (blue)};
\end{tikzpicture}
\end{minipage}\hfill
\begin{minipage}[c]{0.48\textwidth}
\centering
\begin{tikzpicture}[x=0.93cm,y=0.93cm]
\draw[->] (0,0) -- (6.35,0);
\draw[->] (0,0) -- (0,3.4);
\foreach \x/\t in {0/0,1.5/0.25,3/0.5,4.5/0.75,6/1}
 \node[font=\scriptsize,below] at (\x,0) {$\t$};
\foreach \y/\t in {0/0,1.5/1,3/2}
 \node[font=\scriptsize,left] at (0,\y) {$\t$};
\draw[thick,blue,domain=0:1,samples=70,smooth,variable=\t]
 plot ({6*\t},{3*sqrt(max(0,1-\t*\t))});
\draw[thick,orange,domain=0:1,samples=70,smooth,variable=\t]
 plot ({6*\t},{3*\t});
\node[font=\small] at (3,-0.55) {$|n_y|$};
\node[font=\small,rotate=90] at (-0.55,1.5) {Block capacity};
\node[font=\small] at (3,3.45) {(b) Constitutive allocation};
\draw[thick,blue] (0.4,2.75)--(0.9,2.75);
\node[font=\scriptsize,anchor=west] at (1,2.75) {$D_{\mathrm{cap,cent}}$};
\draw[thick,orange] (3.5,2.75)--(4,2.75);
\node[font=\scriptsize,anchor=west] at (4.1,2.75) {$C_{\mathrm{cap,cent}}$};
\end{tikzpicture}
\end{minipage}
\caption{Spectral susceptibility requirements for the
ideal reciprocal, lossless BI polarization compensator.
(a) Minimum total budget over the tunable trajectory,
$B_{\min}^{\mathrm{BI}}(\Theta,|n_y|)
=2\tan(\Theta/4)\max\{\sqrt{1-n_y^2},|n_y|\}$.
White isolines mark budgets $0.5$, $1$, and $1.5$;
the color scale runs from $0$ (white) to $2$ (blue).
(b) At $\Theta=\pi$, separate centered direct electric--magnetic
(blue) and magnetoelectric (orange) peak block capacities
versus $|n_y|$. Their upper envelope is the sharp total
endpoint capacity, which ranges from $\sqrt2$ to $2$.
The two curves represent allocations in a common
optimal centered realization, not independently minimized budgets.
Neither panel represents a control voltage.}
\label{fig:unitary-capacity-angular}
\end{figure*}
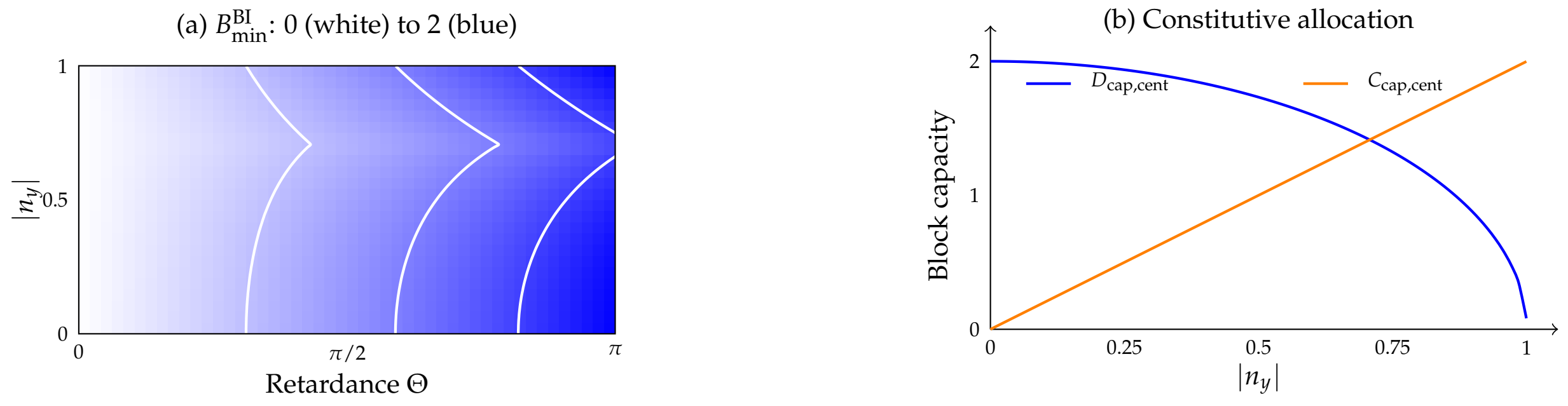

Fig.~\ref{fig:unitary-capacity-angular}
shows the full susceptibility trajectory versus
retardance and $|n_y|$, followed by the direct
and magnetoelectric block capacities at the
half-wave endpoint. Rotation invariance of the
spectral norm removes the remaining azimuthal
dependence, so a two-dimensional trajectory plot
captures the relevant variation without a
redundant spherical-angle map.

The supremum of pointwise minima gives a necessary
peak capacity for any tunable device that covers
the stated transformation family.
The centered construction also supplies a
continuous, ideal sheet-level attaining path,
but it does not establish that a particular bias
mechanism can follow that path. Voltage range,
bias-dependent constitutive constraints, loss,
and dispersion remain additional implementation
requirements.

\subsection{Nonunitary polarization transformations}
\label{sec:nonunitary}

A nonunitary Jones matrix describes a transmitted
polarization response that does not preserve
the norm and inner products of all incident states.
It can attenuate different polarization modes
unequally and change their relative distinguishability,
rather than merely rotate or retard them as a
unitary matrix does.  The transmitted Jones
operator is consequently a contraction, while the
reflected response is available to complete the
lossless scattering process. Our concern is the
minimum susceptibility required by a prescribed
nonunitary transformation, after optimizing these
unconstrained scattering degrees of freedom.

\subsubsection{Polarization-state separation and discrimination}
\label{sec:nonunitary-applications}
\label{sec:nonunitary-separation}
Polarization-state separation provides a
concrete application. Two initially nonorthogonal
states are transmitted as a more distinguishable
pair, at the expense of a rejected or inconclusive
reflection channel. This probabilistic operation
underlies state discrimination
\cite{cheflesBarnett1998separation,chefles2002transformations}
and motivates optical polarization receivers
\cite{clarke2001usd,mohseni2004,gomez2022frio}.

\begin{figure*}[!t]
\centering
\setlength{\tabcolsep}{3pt}
\begin{tabular}{@{}ccc@{}}
\resizebox{.318\textwidth}{!}{
\begin{tikzpicture}[x=1cm,y=1cm,>=stealth,baseline=(current bounding box.south)]
  \path[use as bounding box] (0,0) rectangle (5.65,2.42);
  \tikzset{
    inputstate/.style={->,draw=blue!75!black,line width=.75pt},
    outputstate/.style={->,draw=orange!85!black,line width=.75pt},
    baseaxis/.style={draw=gray!40,densely dashed,line width=.4pt},
    sheet/.style={draw=blue!60!black,fill=blue!12,line width=.5pt}
  }
  \foreach \yy/\delta in {1.74/0,.58/26}{
    \draw[baseaxis] (.30,\yy)--(1.60,\yy);
    \draw[baseaxis] (3.25,\yy)--(4.85,\yy);
    \draw[inputstate] (.45,\yy)--++(22:.63);
    \draw[inputstate] (.45,\yy)--++(-22:.63);
    \draw[outputstate] (3.50,\yy)--++({\delta+36}:.72);
    \draw[outputstate] (3.50,\yy)--++({\delta-36}:.72);
    \draw[->,gray!75,line width=.65pt] (1.39,\yy)--(3.19,\yy);
    \draw[sheet] (2.20,\yy-.31) rectangle (2.34,\yy+.31);
    \draw[->,gray!65,densely dashed,line width=.45pt]
        (2.18,\yy-.12)--(1.77,\yy-.36);
  }
  \node[anchor=west,font=\scriptsize\bfseries] at (.18,2.26) {Aligned};
  \node[anchor=east,font=\scriptsize] at (5.45,2.26) {$\delta=0$};
  \node[anchor=west,font=\scriptsize\bfseries] at (.18,1.09) {Rotated};
  \node[anchor=east,font=\scriptsize] at (5.45,1.09) {$\delta\neq0$};
  \draw[orange!75!black,densely dashed,line width=.4pt]
    (3.50,.58)--++(26:.84);
  \node[font=\tiny,anchor=west,text=orange!80!black]
    at (4.25,.78) {$\delta$};
  \node[font=\tiny,anchor=west] at (.18,.10) {Same $a$, $b$ and BI efficiency};
\end{tikzpicture}} &
\includegraphics[width=.318\textwidth]{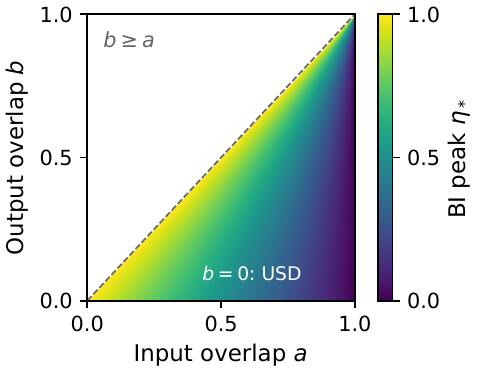} &
\includegraphics[width=.318\textwidth]{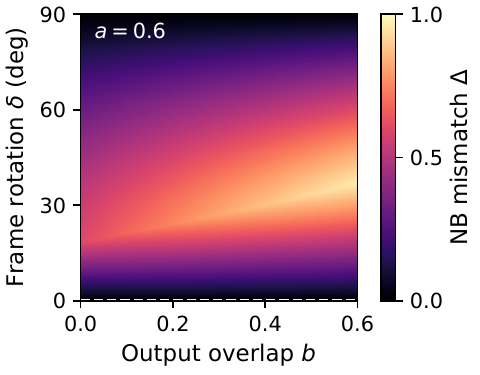} \\
\footnotesize (a) & \footnotesize (b) & \footnotesize (c) \\[3pt]
\includegraphics[width=.318\textwidth]{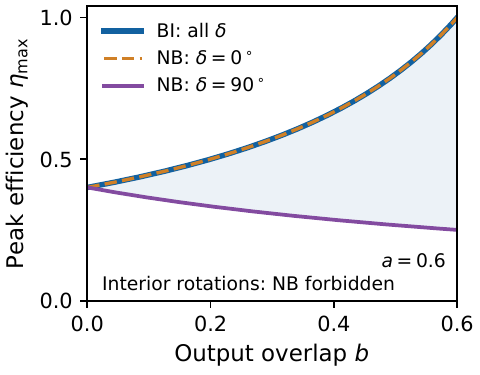} &
\includegraphics[width=.318\textwidth]{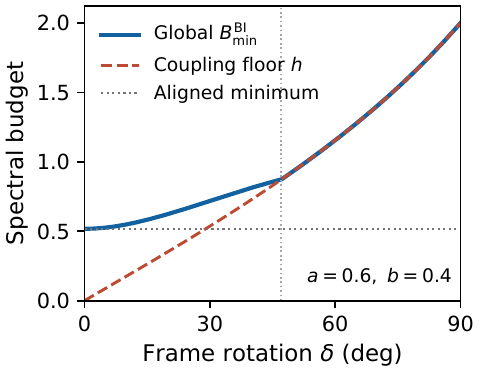} &
\includegraphics[width=.318\textwidth]{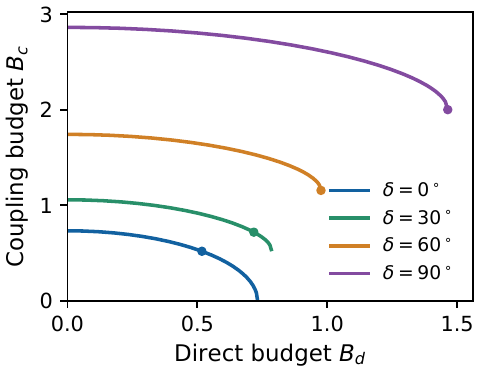} \\
\footnotesize (d) & \footnotesize (e) & \footnotesize (f)
\end{tabular}
\caption{Polarization-state separation and susceptibility requirements.
(a) Aligned and rotated output frames for the same input pair;
blue and orange arrows denote incident and transmitted Jones states,
and dashed arrows denote reflection.
(b) BI maximum transmission efficiency versus the input and output
overlaps $a$ and $b$; white denotes $b\ge a$, outside the separation domain.
(c) NB reciprocity mismatch $\Delta$ at $a=0.6$; zero-mismatch orientations
admit NB transmission.
(d) BI and NB efficiency limits versus target-frame rotation.
(e) Minimum BI susceptibility and magnetoelectric coupling floor for
$a=0.6$, $b=0.4$, at $\eta_*=2/3$.
(f) Globally sharp direct--coupling Pareto frontiers for the same parameters;
markers indicate minimum-budget solutions.}
\label{fig:separation-overview}
\end{figure*}
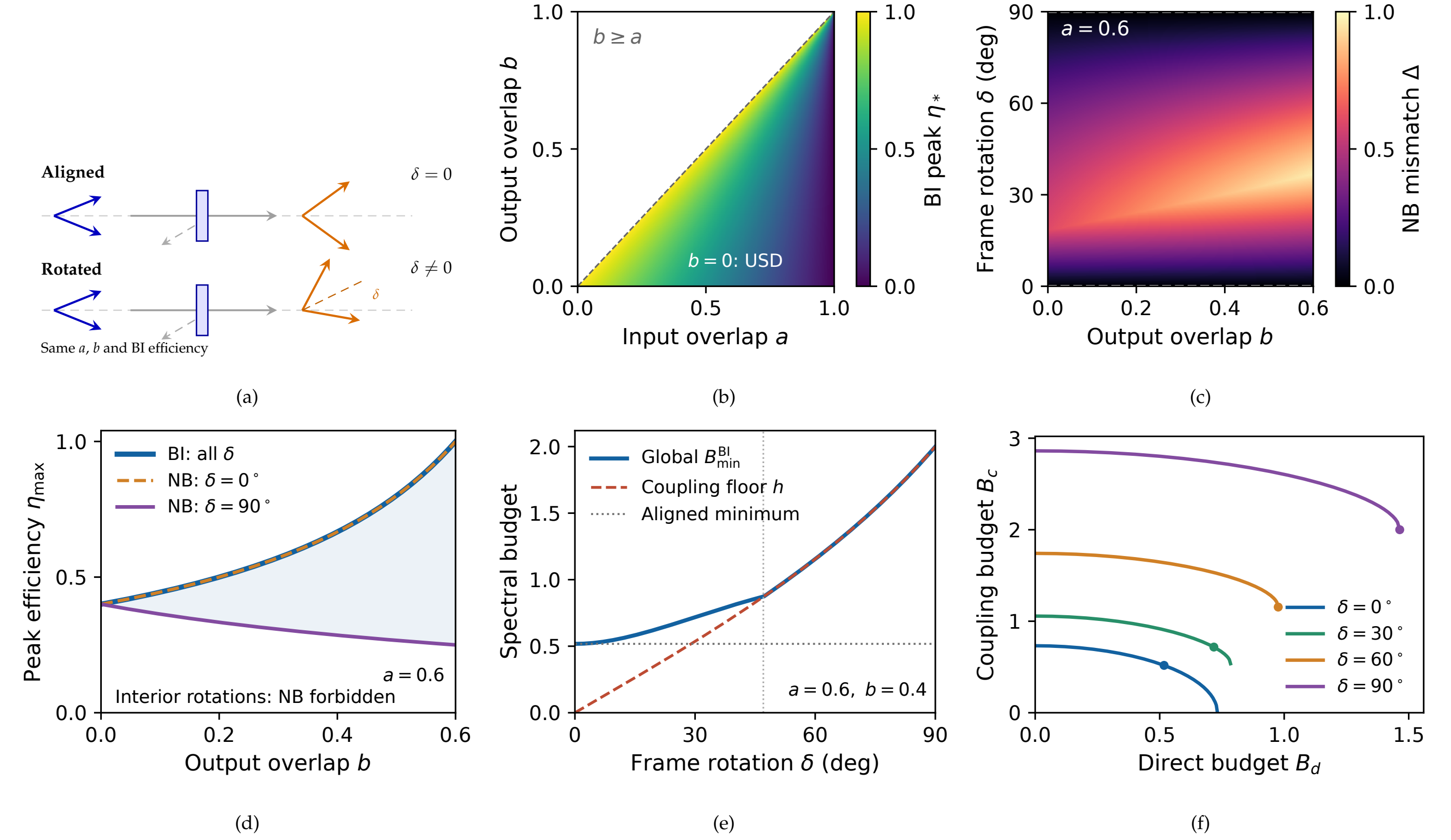

The application specifies both \emph{separation}
and \emph{frame orientation}. The input and output
overlaps $a=|\bm u_+^\dagger\bm u_-|$ and
$b=|\bm v_+^\dagger\bm v_-|$ satisfy $b<a$.
For the symmetric real-Jones states considered
below, aligned input and output pairs share a
symmetry axis, whereas rotating both outputs
through $\delta$ changes their frame orientation
without changing $b$. Figure~\ref{fig:separation-overview}(a)
shows the two geometries. Their BI peak efficiency
is the same established value
$\eta_*=(1-a)/(1-b)$
[Eq.~\eqref{eq:equal-efficiency-limit},
Fig.~\ref{fig:separation-overview}(b)].
This known probability bound determines the
operating point; it does not determine the
minimum susceptibility.

The distinction matters because NB symmetry
restricts the allowed output frames. In this
rotation family, the NB reciprocity mismatch
vanishes only at $\delta=0,\pi/2$
[Fig.~\ref{fig:separation-overview}(c)].
NB reaches the BI efficiency peak when aligned;
at $\delta=\pi/2$ its peak is lower for $b>0$,
and intermediate rotations have no
positive-efficiency NB solution
[Fig.~\ref{fig:separation-overview}(d)].
At $b=0$ the two endpoint peaks coincide.
The general condition is in
\href{supplement.pdf\#section.2}{Supplement~1, Sec.~S2}.
We therefore compare class budgets at a
\emph{common attainable efficiency} for
aligned separation, including the USD limit.
We then hold the BI peak fixed to isolate
the additional orientation-dependent
susceptibility cost
[Fig.~\ref{fig:separation-overview}(e,f)].

\paragraph{Aligned polarization-state separation.}
For aligned symmetric input and output
frames, the state-separation task is
\[
 \begin{gathered}
 \bm u_\pm=
 \begin{pmatrix}\sqrt{(1+a)/2}\\
 \pm\sqrt{(1-a)/2}\end{pmatrix},\\
 \bm v_\pm=
 \begin{pmatrix}\sqrt{(1+b)/2}\\
 \pm\sqrt{(1-b)/2}\end{pmatrix},
 \quad 0\le b<a<1.
 \end{gathered}
\]
At $\eta_*=(1-a)/(1-b)$, an attaining
transmission is $\mathbf T_*=\operatorname{diag}(s,1)$,
where
\[
 s=\sqrt{\frac{(1-a)(1+b)}
 {(1+a)(1-b)}}.
\]
The diagonal transmission is an attaining
witness, but the material minimum must hold
over all admissible transmission phases and
lossless reflection completions. The
phase-global lower bound
(\href{supplement.pdf\#section.5}{Supplement~1, Sec.~S5})
and its attaining completion in
\href{supplement.pdf\#section.4}{Supplement~1, Sec.~S4}
give the sharp frontier
\begin{equation}
 \begin{gathered}
 g(a,b)=2\sqrt{\frac{1-s}{1+s}},\qquad
 B_d^2+B_c^2\ge g(a,b)^2,\\
 B_{\min}^{\mathrm{BI}}(a,b)
 =\frac{g(a,b)}{\sqrt2},\qquad
 B_{\min}^{\mathrm{NB}}(a,b)=g(a,b).
 \end{gathered}
 \label{eq:separation-symmetric-budget}
\end{equation}
The lower quarter-circle in the $(B_d,B_c)$
plane is attainable and anchors the aligned
frontier of Fig.~\ref{fig:separation-overview}(f).
BI can share the constitutive burden between
direct electric--magnetic response and
magnetoelectric coupling; NB must supply
it without coupling. Hence BI requires only
$1/\sqrt2$ of the minimum NB spectral budget,
a sharp $29.3\%$ saving at the \emph{same}
peak transmission efficiency. This is a
result for the specified symmetric frames,
not a general overlap-only cost formula.

\paragraph{Unambiguous discrimination as the orthogonal-output limit.}
\label{sec:nonunitary-discrimination}
The orthogonal-output limit $b=0$ makes the
device interpretation especially direct:
transmitted states can be distinguished
without error, while reflection supplies
the inconclusive outcome required for
unambiguous state discrimination (USD).
The known equal-prior success ceiling is
$\eta_*=1-a$
\cite{ivanovic1981,dieks1982,peres1982,barnett2009};
see also polarization-encoded communication
and optical USD experiments
\cite{bennett1992b92,clarke2001usd,mohseni2004}.
The aligned symmetric example is
\begin{equation}
 \begin{gathered}
 \bm u_{\pm}=
 \begin{pmatrix}
 \sqrt{(1+a)/2}\\
 \pm\sqrt{(1-a)/2}
 \end{pmatrix},\qquad
 \bm v_{\pm}=\frac1{\sqrt2}
 \begin{pmatrix}1\\ \pm1\end{pmatrix},\\
 \eta_*=1-a,\qquad
 \mathbf T_{\mathrm{diag}}
 =\operatorname{diag}(s,1),\quad
 s=\sqrt{\frac{1-a}{1+a}}.
 \end{gathered}
 \label{eq:discrimination-fixture}
\end{equation}
Unlike the nonorthogonal-output case,
$b=0$ leaves the relative transmitted phase
unconstrained at the efficiency ceiling.
A diagonal witness alone therefore does
not establish the global material minimum.
Optimizing the full phase family
(\href{supplement.pdf\#section.5}{Supplement~1, Sec.~S5})
and constructing an attaining completion
(\href{supplement.pdf\#section.4}{Supplement~1, Sec.~S4})
give the sharp budgets
\begin{equation}
 \begin{aligned}
 B_{\min}^{\mathrm{BI}}(a)
 &=\sqrt{\frac{2a}{1+\sqrt{1-a^2}}},\\
 B_{\min}^{\mathrm{NB}}(a)
 &=2\sqrt{\frac{a}{1+\sqrt{1-a^2}}},
 \qquad
 \frac{B_{\min}^{\mathrm{NB}}}
      {B_{\min}^{\mathrm{BI}}}=\sqrt2 .
 \end{aligned}
 \label{eq:discrimination-budget}
\end{equation}
Both classes reach the USD ceiling for
these frames; BI saves $29.3\%$ in budget.
Rotated orthogonal outputs also realize USD, but
our closed-form susceptibility bounds here apply
only to the aligned frames.

\paragraph{Rotated polarization-state separation.}
\label{sec:nonunitary-rotated}
To isolate the effect of orientation, rotate
both targets by the same Jones-plane angle:
$\bm v_\pm(\delta)=R(\delta)\bm v_\pm$,
with $0<b<a<1$ and $0\le\delta\le\pi/2$.
The pair's mutual overlap and BI peak success
probability do not change. Yet the
peak-efficiency transmission representative
becomes $R(\delta)\operatorname{diag}(s,1)$,
up to a common phase, rather than the
aligned diagonal matrix. Its off-diagonal
entries are unequal for nonzero $\delta$;
this nonsymmetric part cannot be supplied
by NB at the BI peak. BI can instead
accommodate the change through
magnetoelectric response. Frame orientation
therefore constitutes a material-resource
variable independent of the overlap magnitudes.

With $\kappa=\cos\delta$ and
$\rho=\sqrt{(1-s)/(1+s)}$, define
\begin{equation}
 m=\frac{4\rho}{1+\kappa},\qquad h=2\tan(\delta/2).
 \label{eq:separation-rotation-scales}
\end{equation}
The sharp bound $B_c\ge h$ holds at
the BI efficiency ceiling, even after
optimizing the free transmission phase
and reflection completion. In this
symmetric-frame family, $h$ depends only
on the target rotation, not on the input
or output overlaps. Thus nonzero rotation
creates an unavoidable magnetoelectric
coupling requirement rather than merely
making coupling a cheaper design option.
Figure~\ref{fig:separation-overview}(e)
compares this floor with the minimum BI
budget for $a=0.6$, $b=0.4$. This
overlap-independent floor is not asserted
for arbitrary Jones frames.

For $m\le2$, the globally sharp
direct--coupling Pareto frontier,
parameterized by $0\le v\le1$, is
\begin{equation}
 \begin{aligned}
 B_d&=mv,\\
 B_c&=\frac{\sqrt{m^2(1-v^2)+4h^2}+m\sqrt{1-v^2}}2.
 \end{aligned}
 \label{eq:separation-rotated-frontier}
\end{equation}
The sufficient condition $s\ge3/5$
ensures $m\le2$ throughout the rotation
range, including the plotted benchmark.
Figure~\ref{fig:separation-overview}(f)
shows how the direct--coupling Pareto
frontier moves away from the aligned
quarter-circle ($h=0$); its marked points
identify minimum total budgets. Two
physical regimes emerge: for $m>h$,
the optimum balances direct and coupling
resources, whereas for $m\le h$ the
orientation-imposed coupling floor
determines the minimum. Thus the frame
angle changes not only the required
susceptibility, but which constitutive
block limits the design. Writing
$\mathcal R=m^4+4m^2h^2-4h^4$ gives
\begin{equation}
 B_{\min}^{\mathrm{BI}}(a,b,\delta)=
 \begin{cases}
 h,&m\le h,\\[2pt]
 \displaystyle\frac12\sqrt{m^2+2h^2+\sqrt{\mathcal R}},&m>h.
 \end{cases}
 \label{eq:separation-rotation-sharp}
\end{equation}
Panels (e,f) therefore show an
orientation-driven change in both the minimum
susceptibility and its limiting constitutive block,
at fixed overlaps and BI efficiency.
The frontier is globally sharp for $m\le2$;
the scalar minimum is also sharp when its
value is at most two. Elsewhere noncentered
completions may improve the cost
(\href{supplement.pdf\#section.6}{Supplement~1, Sec.~S6}).
Rotated $b=0$ tasks lie outside the stated
two-phase frontier formulas.

\section{Conclusions}
\label{sec:conclusions}

We have established globally attained minimum
effective-susceptibility bounds for prescribed
polarization transformations in reciprocal, lossless
dipolar metasurfaces. These bounds quantify the
least constitutive response required to perform a
given electromagnetic function. Magnetoelectric
coupling can substantially reduce this requirement,
while offering no resource advantage for certain
transformations. For nonunitary polarization-state
separation, the sharp resource frontiers further
reveal how the optimal allocation between direct
electric--magnetic and magnetoelectric responses
depends on the prescribed transformation, including
transitions between balanced and coupling-limited regimes.

The dimensionless susceptibility budget provides a
quantitative benchmark for inverse design. The
analytical minimum identifies a lowest-cost ideal
effective sheet and, where available, its optimal
phase and scattering-completion choices. These
provide targets and lower-bound benchmarks for
numerical meta-atom inverse design
\cite{molesky2018,pearsonHum2022,naseri2022,
wang2024polarizers}, including physics-informed
geometry-to-susceptibility models
\cite{park2026gstcneural}. The gap between the
retrieved effective-susceptibility budget of a
fabricated design and its ideal minimum quantifies
the additional constitutive resources associated
with that architecture and its constraints.
Such comparisons require physically consistent
effective-susceptibility retrieval rather than
treating arbitrary latent design parameters as
constitutive tensors.

Extensions to passive absorption
\cite{guan2024,achouri2021}, higher multipolar GSTCs
\cite{achouriMartinPRB2021,tiukuvaara2023quadrupolar,
achouriTiukuvaaraMartin2023}, and oblique incidence
\cite{achouri2021,abdelrahmanMonticone2023}
would broaden the resource-optimization framework
to dissipative responses, additional constitutive
channels, and angle-dependent electromagnetic
interactions. These extensions require appropriately
defined resource measures, particularly when comparing
reactive and dissipative responses or multipoles of
different dimensions. The relationship between
susceptibility-optimal Jones operators and nonorthogonal
eigenpolarizations \cite{yuan2024} also merits
investigation. A further direction is continuously
tunable polarization conversion: along prescribed
trajectories on the Poincar\'e sphere, pointwise
susceptibility minima need not form a continuous,
control-accessible path. Minimizing the peak budget
subject to continuous sheet parameters and realizable
bias-dependent responses would connect these bounds
to reconfigurable devices. Practical implementations
must also account for material loss, dispersion,
bandwidth, substrate effects, and fabrication constraints.

\section*{Disclosures}
The authors declare no conflicts of interest.

\section*{Data Availability}
No experimental data were generated in this study.
The Lean 4 formalization of selected theoretical results
was developed with LLM assistance. The Lean 4 verification code and theorem statements are available at
\url{https://github.com/rshegde/gstc-arbpol}.

\bibliography{references}
\end{document}


\maketitle

The sections follow the order of the main article: GSTC recovery
and global minima (Secs.~S1--S2), unitary transformations (Sec.~S3),
and polarization-state separation in aligned, orthogonal-output,
and rotated frames (Secs.~S4--S6).

\section{GSTC recovery and polynomial feasibility}
\label{app:gstc-derivation}
\label{app:minimum-derivation}

The GSTC forward and inverse scattering relations are standard
\cite{pfeiffer2014,achouri2015general,achouri2021,
szymanski2021bianisotropic}. We record the field normalization
needed for the spectral budgets and establish the exact
polynomial conditions used in Sec.~2B.
With $\mathbf J=\left(\begin{smallmatrix}0&-1\\1&0\end{smallmatrix}\right)$,
set
\begin{equation}
\begin{gathered}
\mathbf F^{-1}=
\begin{pmatrix}
Z_0^{-1/2}\mathbf I_2&Z_0^{1/2}\mathbf J\\
Z_0^{-1/2}\mathbf I_2&-Z_0^{1/2}\mathbf J
\end{pmatrix},\\
\mathbf H_{\mathbf S}
=-\frac{1}{2\omega}\mathbf F^{-\mathsf T}
\mathbf K_{\mathbf S}\mathbf F^{-1}.
\end{gathered}
\label{eq:app-field-map}
\end{equation}
The BI susceptibility blocks are recovered as
\begin{equation}
\mathbf A=\operatorname{Re}(\mathbf H_{\mathbf S})_{11}/\epsilon_0,\quad
\mathbf D=\operatorname{Re}(\mathbf H_{\mathbf S})_{22}/\mu_0,\quad
\mathbf G=c_0\operatorname{Im}(\mathbf H_{\mathbf S})_{12}.
\label{eq:brec-inverse-budget-map}
\end{equation}
For $\mathbf L=2\mathbf I_4-\jj\mathbf K(\bm p)$,
the standard Cayley relation gives
$\mathbf S=\bm\Theta\mathbf L^{-1}(2\mathbf I_4+\jj\mathbf K)$.
Hermiticity of $\mathbf K$ makes $\mathbf L$ invertible for
every finite real $\bm p$. Setting $q_T=\det\mathbf L$
and $\mathbf Q_T=[\bm\Theta\operatorname{adj}(\mathbf L)
(2\mathbf I_4+\jj\mathbf K)]_{21}$ yields the
polynomial identity $q_T\mathbf T=\mathbf Q_T$ with $q_T\ne0$.
For any normalized target $\bm v$, the two conditions
$\det[\bm v,\mathbf Q_T\bm u]=0$ and
$\|\mathbf Q_T\bm u\|^2=\eta|q_T|^2$
are therefore necessary and sufficient for
$\mathbf T\bm u=\alpha\bm v$, $|\alpha|^2=\eta$.
Applied to both inputs, they give precisely
Eq.~\textup{(16)}, including zero efficiency
and dependent targets, without imposing zero reflection.

\section{Global minima and scattering completions}
\label{app:general-results}

This section supplies the attainment argument and the complete
BI and NB scattering parameterizations underlying the general
minimum-susceptibility results of Sec.~2D.

\paragraph{Attainment.}
The feasible set \textup{(17)} is
closed, and spectral-budget sublevel sets are
compact, so the minimum is attained whenever
the feasible set is nonempty. Principal-minor
inequalities express the spectral caps; for
algebraic data, the minimum is algebraic by
real-algebraic projection.

\paragraph{BI scattering completion and recovery.}
For independent input and target frames
$\mathbf U=[\bm u_1,\bm u_2]$ and
$\mathbf V=[\bm v_1,\bm v_2]$, the two prescribed
conversions at common efficiency $\eta$ require
\begin{equation}
\mathbf X=e^{\jj\phi}\sqrt{\eta}\,
\mathbf V\operatorname{diag}(1,e^{\jj\psi})\mathbf U^{-1},
\qquad
\Psi_\eta=\{\psi:|d-\eta e^{\jj\psi}c|\le1-\eta\}.
\label{eq:brec-transmission-family}
\end{equation}
Here $\Psi_\eta$ enforces contraction and $\phi$
is free. For $a,b>0$, the Gram bound fixes
$e^{\jj\psi}=d\overline c/(ab)$ at the BI peak.
Choose the common-phase-free transmission
$\mathbf X_0$ with singular values $(1,t)$,
$0<t\le1$. When $t<1$, its power defect has rank one:
$\mathbf I_2-\mathbf X_0^{\mathsf T}\overline{\mathbf X_0}
=(1-t^2)\bm n\bm n^\dagger$, $\|\bm n\|=1$.
Consequently every symmetric-unitary scattering
completion with this transmission block is
$\mathbf S=e^{\jj\phi}
\left(\begin{smallmatrix}\mathbf Z&\mathbf X_0^{\mathsf T}\\
\mathbf X_0&\mathbf Y\end{smallmatrix}\right)$, with
\begin{equation}
\begin{aligned}
\mathbf Z&=e^{\jj\theta}\sqrt{1-t^2}\,\bm n\bm n^{\mathsf T},\\
\mathbf Y&=-(\overline{\mathbf X_0^{\mathsf T}})^{-1}
\overline{\mathbf Z}\mathbf X_0^{\mathsf T}.
\end{aligned}
\label{eq:app-brec-completion}
\end{equation}
Unitarity fixes $\mathbf Z$ up to $\theta$ and
then determines $\mathbf Y$, making the family
complete. At $t=1$ both reflection blocks vanish.
The generic nonunitary case therefore leaves only
$\phi$ and $\theta$. Retain completions satisfying
$\det(\bm\Theta\mathbf S+\mathbf I_4)\ne0$.
Inverse Cayley \textup{(12)}
and block recovery \eqref{eq:brec-inverse-budget-map}
then yield the minimum \textup{(21)}.

Dependent target frames can give singular
transmission, outside the inverse formula for
$\mathbf Y$. Symmetric-unitary completions still
exist and the peak is covered by
\textup{(10)}
\cite{chefles2002transformations,cheflesJozsaWinter2004};
\textup{(21)} applies to
nonsingular frames.

\paragraph{NB parity completion and recovery.}
At positive common efficiency and nonzero
overlaps $a=|d|$, $b=|c|$, the NB phase
condition fixes
$\Lambda=h_{12}/h_{21}$ whenever both
cross-frame products are nonzero.
If $|\Lambda|=1$, the BI and NB peak
efficiencies for state separation agree
precisely when the NB phase equals the
phase selected by the BI passivity bound:

\begin{equation}
 \eta_{\max}^{\mathrm{NB}}
 =\eta_{\max}^{\mathrm{BI}}
 \quad\Longleftrightarrow\quad
 \frac{\Lambda c}{d}=\frac{b}{a},
 \qquad a,b>0,\quad|\Lambda|=1.
 \label{eq:nonunitary-peak-equality}
\end{equation}
Here $d=\bm u_1^\dagger\bm u_2$,
$c=\bm v_1^\dagger\bm v_2$, and
$\eta_{\max}^{\mathrm{BI}}=(1-a)/(1-b)$
for $0<b<a<1$. Vanishing overlaps make the
passive ceiling independent of the relative
phase, while unequal cross-frame magnitudes
exclude a positive-efficiency NB representative.

Using the cross-frame bilinear products $h_{12}$ and
$h_{21}$ defined in Sec.~2B,
NB transmission symmetry imposes
\begin{equation}
\alpha_1 h_{12}=\alpha_2 h_{21}.
\label{eq:jones-nb-cross-condition}
\end{equation}
Every complex symmetric contraction is attainable
by a finite NB sheet up to a common phase:
\begin{equation}
\mathbf X^{\mathsf T}=\mathbf X,\quad
\|\mathbf X\|_2\le1
\quad\Longrightarrow\quad
\exists\,\bm p,\zeta:\ 
\mathbf G(\bm p)=\mathbf0,\ 
|\zeta|=1,\ 
\mathbf T(\bm p)=\zeta\mathbf X.
\label{eq:nb-symmetric-contraction-reachability}
\end{equation}
NB scattering is equivalent to symmetric-unitary
parity blocks $\mathbf X,\mathbf Y$ with
$\mathbf T=(\mathbf X+\mathbf Y)/2$.
Takagi factorization \cite{hornZhang2012takagi}
guarantees such pairs even for singular symmetric
contractions. The minimum ranges over \emph{all}
pairs with the prescribed midpoint.
For a pole-free pair and common phase $|z|=1$,
inverse Cayley gives
\begin{equation}
\begin{aligned}
\mathbf A(z,\mathbf X)
&=\frac{2}{\jj\omega\epsilon_0Z_0}
(\mathbf I_2-z\mathbf X)(\mathbf I_2+z\mathbf X)^{-1},\\
\mathbf D(z,\mathbf Y)
&=-\frac{2Z_0}{\jj\omega\mu_0}
\mathbf J(\mathbf I_2-z\mathbf Y)
(\mathbf I_2+z\mathbf Y)^{-1}\mathbf J .
\end{aligned}
\label{eq:em-inverse-budget-map}
\end{equation}
Optimizing the recovered budget over every
pole-free parity pair and common phase gives
the global NB minimum \textup{(22)}.

\section{Unitary polarization transformations: sharpness and exact minima}
\label{app:orthogonal}
\label{app:specialized-results}

These calculations support the phase-optimized susceptibility bounds
and the linear-to-elliptical and tunable applications of
Sec.~3A.

\subsection{Phase centering and global susceptibility bounds}
\label{app:unitary-bounds}

For orthonormal Jones frames let
$e^{2\jj\varphi_0}=\det(\mathbf V\mathbf U^\dagger)$ and
\begin{equation}
\begin{gathered}
\mathbf M(t)=e^{-\jj\varphi_0}\mathbf V
\operatorname{diag}(e^{\jj t},e^{-\jj t})
\mathbf U^\dagger
=q(t)\mathbf I_2+\jj\bm r(t)\cdot\bm\sigma,\\
(q(t),\bm r(t))=(q_c,\bm a)\cos t+
(q_s,\bm b)\sin t .
\end{gathered}
\label{eq:orth-su2}
\end{equation}
The coefficient four-vectors are orthonormal and
$q^2+|\bm r|^2=1$.
For the traceless inverse-Cayley Pauli block with
coordinates $(x,y,z)$, the recovered dimensionless
susceptibilities are
\begin{equation}
\begin{aligned}
k_0\mathbf A&=-e
\begin{pmatrix}z&x\\x&-z\end{pmatrix},&
k_0\mathbf D&=m
\begin{pmatrix}z&x\\x&-z\end{pmatrix},\\
k_0\mathbf G&=vy\,\mathbf I_2 .
\end{aligned}
\label{eq:orth-material}
\end{equation}
Here $e=k_0/(\omega Z\epsilon_0)$,
$m=k_0Z/(\omega\mu_0)$, and $v=k_0c_0/\omega$,
all unity in matched free space.

For $\mathbf M=q\mathbf I_2+
\jj\sin\theta\,\hat{\bm n}\cdot\bm\sigma$,
every pole-free common phase yields
\begin{equation}
\mathbf K(\alpha)=u\mathbf I_2+
h\hat{\bm n}\cdot\bm\sigma,\quad
u=\frac{2\sin\alpha}{\cos\alpha+q},\quad
h=\frac{2\sin\theta}{\cos\alpha+q}.
\label{eq:app-uncentred}
\end{equation}
The direct-block spectral norm is
$|u|+|h|\sqrt{n_x^2+n_z^2}$.
Since $|\cos\alpha+q|\le1+|q|$, the
spectral budget is at least
$2|\sin\theta|
\max\{w\sqrt{n_x^2+n_z^2},v|n_y|\}/(1+|q|)$.
The pole-free centered phase $\alpha=0$
for $q\ge0$ or $\alpha=\pi$ otherwise
attains equality. This proves
Eq.~\textup{(23)} over all phases.
NB additionally requires $r_y(t)=0$.
For nonorthogonal unitary tasks ($0<a=b<1$),
the fixed relative phase
\textup{(28)}
and the same common-phase bound prove
Eq.~\textup{(29)}.
The centered optimum belongs to NB exactly
when the transmission is symmetric.

At any NB-admissible $t_0$, unitarity gives
\begin{equation}
B_{\min}^{\mathrm{NB}}
\le2w\frac{|\bm r(t_0)|}{1+|q(t_0)|}
=2w\sqrt{\frac{1-|q(t_0)|}{1+|q(t_0)|}}
\le2w .
\label{eq:app-delta-bound}
\end{equation}
Class inclusion gives $\Delta B\le2w$.
Sharpness follows from the orthonormal family
\begin{equation}
(q_c,\bm a)=(0,1,0,0),\quad
(q_s,\bm b)=(\sqrt{1-2\delta^2},0,\delta,\delta),
\quad 0<\delta<1/\sqrt2 .
\label{eq:app-family}
\end{equation}
Its only NB phases are $t=0,\pi$,
with $B_{\min}^{\mathrm{NB}}=2w$;
at $t=\pi/2$, BI has budget at most
$2\max(w,v)\delta/(1+\sqrt{1-2\delta^2})\to0$.
Explicit Jones representatives for this family are
$\mathbf M_0=\jj\sigma_x$ and
$\mathbf M_1=\sqrt{1-2\delta^2}\mathbf I_2+
\jj\delta(\sigma_y+\sigma_z)$:
$\mathbf M_0^{-1}\mathbf M_1$ is unitarily
conjugate to $\jj\sigma_z$.
Thus the absolute savings approach $2w$
and the fractional savings approach unity.
The near-identity discontinuity concerns
the optimized budget, not the fields.

\subsection{Linear-to-elliptical conversion}
For the linear-rotation family in
Sec.~3A.1, take
$\mathbf T=R(\theta)
\operatorname{diag}(e^{\jj t},e^{-\jj t})$.
Its determinant-centered Pauli coordinates are
$q=\cos\theta\cos t$,
$r_x=\sin\theta\sin t$,
$r_y=-\sin\theta\cos t$, and
$r_z=\cos\theta\sin t$.
Consequently, in free space,
\[
B_d(t)=\frac{2|\sin t|}
 {1+|\cos\theta\cos t|},\qquad
B_c(t)=\frac{2|\sin\theta\cos t|}
 {1+|\cos\theta\cos t|}.
\]
For $0<|\theta|<\pi/2$, the only NB-admissible
relative phases have $\cos t=0$, hence
$B_{\min}^{\mathrm{NB}}=2$.
By contrast, choosing
$t=\tan^{-1}|\sin\theta|$ balances both
blocks and yields the global BI minimum
\[
B_{\min}^{\mathrm{BI}}
 =\frac{2|\sin\theta|}
 {\sqrt{1+\sin^2\theta}+|\cos\theta|}
 \longrightarrow0\quad(\theta\to0).
\]
At $\theta=0$, $t=0$ is also NB-admissible
and the transparent sheet gives zero budget.
This proves the specific near-identity cost
discontinuity used in Sec.~3A.1.

\subsection{Tunable polarization transformations}
For the unitary rotation
$\mathbf U(\hat{\bm n},\Theta)
=\cos(\Theta/2)\mathbf I_2-
\jj\sin(\Theta/2)\hat{\bm n}\cdot\bm\sigma$,
consider a noncentered common phase $\alpha$.
The corresponding recovery coefficients are
$u=2\sin\alpha/(\cos\alpha+\cos(\Theta/2))$
and $h=2\sin(\Theta/2)/
(\cos\alpha+\cos(\Theta/2))$.
The direct-block spectral norm contains
$|u|+|h|\sqrt{n_x^2+n_z^2}$, and the coupling
block has norm $|h n_y|$.
Since $|\cos\alpha+\cos(\Theta/2)|
\leq1+\cos(\Theta/2)$, the centered
choice $\alpha=0$ attains the global lower
bound in Eq.~\textup{(33)}.
For
$g(\hat{\bm n})=\max\{\sqrt{1-n_y^2},|n_y|\}$,
\begin{equation}
\frac{\partial B_{\min}^{\mathrm{BI}}}{\partial\Theta}
=\frac{g(\hat{\bm n})}{2}\sec^2(\Theta/4)>0.
\label{eq:rotation-monotonicity}
\end{equation}
This gives the half-wave capacity in
Eq.~\textup{(34)}.

\section{Aligned polarization-state separation: attaining completions}

For the aligned case of Sec.~3B.1,
we construct the lossless completions that attain the sharp
susceptibility frontier, including the orthogonal-output limit.

\label{app:nonunitary-bounds}
\label{app:nonunitary-diagonal}

For aligned symmetric separation,
$\mathbf T_{\mathrm{diag}}=\operatorname{diag}(s,1)$,
$s=\sqrt{(1-a)(1+b)/[(1+a)(1-b)]}$,
$0\le b<a<1$; the USD endpoint is $b=0$.
With $r=\sqrt{1-s^2}$ and
$\mathbf E_{ij}=\bm e_i\bm e_j^{\mathsf T}$,
the symmetric-unitary completion
\[
\mathbf S(\theta)=
\begin{pmatrix}
re^{\jj\theta}\mathbf E_{11}&\mathbf T_{\mathrm{diag}}\\
\mathbf T_{\mathrm{diag}}&-re^{-\jj\theta}\mathbf E_{11}
\end{pmatrix}
\]
is pole-free since
$\det(\bm\Theta\mathbf S+\mathbf I_4)=8(1+s)>0$.
Its inverse-GSTC recovery, with $g_0=2r/(1+s)$,
is
\begin{equation}
\begin{aligned}
k_0\mathbf A&=-g_0\sin\theta\,\mathbf E_{11},\\
k_0\mathbf D&=g_0\sin\theta\,\mathbf E_{22},\\
k_0\mathbf G&=-g_0\cos\theta\,\mathbf E_{12}.
\end{aligned}
\label{eq:app-nonunitary-diagonal-cayley}
\end{equation}
Thus $B_d=g_0|\sin\theta|$ and
$B_c=g_0|\cos\theta|$ trace the entire
quarter-circle $B_d^2+B_c^2=g_0^2$.
The balanced and coupling-free points attain,
respectively, $B_{\mathrm{BI}}=g_0/\sqrt2$
and $B_{\mathrm{NB}}=g_0$.
For $b>0$ Gram saturation fixes the relative
phase. The phase-global certificate for the
orthogonal-output limit $b=0$ appears in
Sec.~\ref{supp:discrimination}.

\section{Unambiguous discrimination: phase-global lower bound}
\label{supp:discrimination}
The attaining construction of Sec.~\ref{app:nonunitary-diagonal}
does not by itself establish a global lower bound when the
output states are orthogonal: their relative transmission phase
is free. We therefore optimize over the complete phase family.

For the symmetric USD fixture \textup{(37)}
put $s=\sqrt{(1-a)/(1+a)}$, $r=\sqrt{1-s^2}$,
$0<a<1$, and $p=e^{\jj\psi}$,
$t=e^{-\jj\phi}$, $q=e^{-\jj\vartheta}$.
The complete symmetric-unitary scattering family has
\begin{equation}
\begin{gathered}
\mathbf X_0=\mathbf W(p)\operatorname{diag}(s,1),
\quad
\mathbf W(p)=\tfrac12
\begin{pmatrix}1+p&1-p\\1-p&1+p\end{pmatrix},\\
\mathbf S=t^{-1}
\begin{pmatrix}
r q^{-1}\mathbf E_{11}&\mathbf X_0^{\mathsf T}\\
\mathbf X_0&-rq\mathbf W(p)\mathbf E_{11}
\mathbf W(p)^{\mathsf T}
\end{pmatrix}.
\end{gathered}
\label{eq:app-discrimination-all-completions}
\end{equation}
Here $\mathbf E_{11}=\operatorname{diag}(1,0)$;
rank-one unitarity fixes the first reflection block up
to a phase, and mixed unitarity fixes the second.
Only pole-free members correspond to finite sheets.

The inverse-Cayley entry
$\gamma=(\mathbf K_{\mathbf S})_{13}
=\tfrac{k_0}{2}(-A_{xx}+D_{yy})-\jj k_0G_{xy}$
obeys $B\ge|\gamma|/\sqrt2$ and
$B_{\mathrm{NB}}\ge|\gamma|$.
Writing its full phase dependence as
\begin{equation}
\begin{gathered}
\gamma=\frac{2\jj qt\sqrt{1-s^2}\,N}{D},
\qquad N=[p(t+2)+t]^2,\\
D=p^2(st^2+2st+t^2+2t+2)\\
\quad+p(2st^3+6st^2+2st+2t^3+2t^2+2t)\\
\quad+2st^3+st^2+2t^4+2t^3+t^2,
\end{gathered}
\label{eq:app-discrimination-gamma}
\end{equation}
the phase-global certificate is the exact
sum-of-squares factorization
\begin{equation}
\begin{aligned}
(1+s)^2|N|^2-|D|^2
=4|p-t^2|^2\bigl(
|p+t|^2|1+t|^2+s|p+2t+1|^2\bigr)\ge0 .
\end{aligned}
\label{eq:app-discrimination-sos}
\end{equation}
Consequently $|\gamma|\ge
g_0=2\sqrt{1-s^2}/(1+s)$ for every phase and
pole-free completion. Since
$B_d\ge|\operatorname{Re}\gamma|$ and
$B_c\ge|\operatorname{Im}\gamma|$, the entire
attainable region satisfies
$B_d^2+B_c^2\ge g_0^2$.
The diagonal completion in
Sec.~\ref{app:nonunitary-diagonal} attains
every point of the nonnegative quarter-circle;
hence its boundary is globally sharp, giving
\begin{equation}
B_{\min}^{\mathrm{BI}}=g_0/\sqrt2,\qquad
B_{\min}^{\mathrm{NB}}=g_0 .
\label{eq:app-discrimination-certified}
\end{equation}
These conclusions concern the specified USD
Jones frames, not arbitrary frame transformations.

\section{Rotated polarization-state separation}
\label{supp:rotated-sharp}
This section supports the orientation-dependent frontier of
Sec.~3B.1. We first parameterize all
lossless completions, then establish the sharp centered regime
and describe the noncentered and orthogonal-output cases.

For $0<b<a<1$, $0\le\delta\le\pi/2$, define
$\eta_*=(1-a)/(1-b)$,
$s=\sqrt{(1-a)(1+b)/[(1+a)(1-b)]}$,
$\rho=\sqrt{(1-s)/(1+s)}$, $r=\sqrt{1-s^2}$,
$c=\cos\delta$, $j=\sin\delta$,
$m=4\rho/(1+c)$, and $h=2j/(1+c)$.
Gram saturation fixes the relative transmission phase.
All reciprocal symmetric-unitary scattering completions are
\begin{equation}
\begin{gathered}
\mathbf X=e^{\jj\phi}\mathbf R\operatorname{diag}(s,1),\quad
\mathbf Z=r e^{\jj\beta}\mathbf E_{11},\\
\mathbf Y=-r e^{\jj(2\phi-\beta)}
\mathbf R\mathbf E_{11}\mathbf R^{\mathsf T},\quad
\mathbf S=\begin{pmatrix}
\mathbf Z&\mathbf X^{\mathsf T}\\
\mathbf X&\mathbf Y
\end{pmatrix},
\end{gathered}
\label{eq:rot-complete}
\end{equation}
where $\mathbf R=\mathbf R(\delta)$ and
$\phi,\beta$ are unrestricted phases.
The rank-one defect fixes $\mathbf Z$, and mixed
unitarity fixes $\mathbf Y$.

Using conjugation symmetry take
$x=\cos\phi$, $y=\sin\phi\ge0$ and
$v=|\cos(\pi/2+\phi-\beta)|$. Set
$F=(1+s)(x+c)^2-(1-s)y^2$ and
\begin{equation}
\begin{aligned}
p_0&=\frac{2y[2x+(1+s)c]}F,&
p_1&=\frac{2(1-s)y}F,\\
t&=\frac{2r(x+c)}F,&u&=\frac{2r(1+cx)}F,\\
w&=\frac{2rjy}F,&q&=\frac{2j(1+s)(x+c)}F.
\end{aligned}
\label{eq:rot-six-scalars}
\end{equation}
The \emph{exact} direct and coupling spectral
resources of the complete family are
\begin{equation}
\begin{aligned}
\mathcal D(x,v)&=\max_{\epsilon=\pm1}
\left\{|p_0+\epsilon tv|
+\sqrt{(p_1+\epsilon uv)^2+w^2(1-v^2)}\right\},\\
\mathcal C(x,v)&=\sqrt{t^2(1-v^2)+q^2}
+\sqrt{t^2(1-v^2)+w^2}.
\end{aligned}
\label{eq:rot-exact-image}
\end{equation}
Here $-1\le x\le1$, $0\le v\le1$,
and $F\ne0$ is precisely the inverse-Cayley
pole exclusion, since
$\det(I+\Theta S)=2e^{2\jj\phi}F$.
Conversely, every such $(x,v)$ is attained by
$\phi=\arccos x$ and
$\beta=\pi/2+\phi-\arccos v$.

\paragraph{Global frontier and its domain.}
At the centered phase $\phi=0$, the attainable arc is
\begin{equation}
D=mv,\qquad
C=\frac{\sqrt{m^2(1-v^2)+4h^2}
+m\sqrt{1-v^2}}{2},\quad 0\le v\le1.
\label{eq:rot-frontier}
\end{equation}
For $F>0$, the exact resource image
\eqref{eq:rot-exact-image} dominates the centered
point with the same $v$:
$2|t|\ge m$, $|q|\ge h$, and
$\mathcal D\ge2|t|v$.
For $F<0$, $|F|\le(1-s)y^2$ instead gives
$D\ge|p_1|\ge2/y\ge2$ and
$C\ge|w|\ge2j/(\rho y)\ge h$.
Consequently the entire centered arc is globally
Pareto optimal provided
\begin{equation}
m\le2\quad\Longleftrightarrow\quad
\rho\le(1+\cos\delta)/2 .
\label{eq:rot-frontier-domain}
\end{equation}
The condition $s\ge3/5$ suffices uniformly in
$\delta$. Its scalar optimum is
\begin{equation}
B_{\mathrm{cent}}=
\begin{cases}
h,&m\le h,\\[2pt]
\frac12\sqrt{m^2+2h^2+
\sqrt{m^4+4m^2h^2-4h^4}},&m>h,
\end{cases}
\label{eq:rot-sharp-min}
\end{equation}
globally sharp whenever $m\le2$, or more
generally when $B_{\mathrm{cent}}\le2$.
These are sufficient conditions, not necessary ones.

\paragraph{Noncentered branch.}
At $\delta=\pi/2$, every $F<0$ completion is
dominated by the attained pair
$(D_*,C_*)=(2\sqrt{1+\rho^{-2}},2/\rho)$:
the lower bounds $|p_1|\ge2$ and
$|w|\ge2/\rho$ combine with
$\mathcal D^2\ge p_1^2+w^2$; equality occurs at
$\phi=\beta=\pi/2$. Thus
\begin{equation}
B_{\min}^{\mathrm{BI}}(a,b,\pi/2)
=\min\{B_{\mathrm{cent}},2\sqrt{1+\rho^{-2}}\}.
\label{eq:rot-right-angle-min}
\end{equation}
For $s=1/10$, $a=149/151$, $b=1/5$,
$B_{\mathrm{cent}}\simeq3.11811$ exceeds
the global minimum $4\sqrt5/3\simeq2.98142$.
Hence phase centering is not universally optimal.

\paragraph{Attainment.}
The centered Pareto arc is attained by the explicit pole-free
completions above. Outside its globally sharp regime, the full
two-phase resource image~\eqref{eq:rot-exact-image} remains
available for numerical minimization. General finite-budget
attainment follows from Sec.~\ref{app:general-results}.

\paragraph{Scope at orthogonal output.}
The formulas in this section assume $0<b<a<1$.
At $b=0$, the relative transmission phase is also free,
so the two-phase completion and its centered-frontier
certificate cannot be applied without further optimization.
The aligned $b=0$ case is treated in Sec.~\ref{supp:discrimination}.